\documentclass[%
  aip,
  pop,   % cambia 'pop' con journal code
  amsmath,amssymb,
  preprint  % preprint -> bozza reprint -> finale
]{revtex4-2}

\usepackage{graphicx}
\usepackage{siunitx}
\usepackage{bm}          % bold math
\usepackage{hyperref}    % links
\usepackage{physics}     % comandi utili
\usepackage{cleveref}    % riferimenti intelligenti
\usepackage{enumitem}
\usepackage{multirow}
\usepackage{tikz}
\allowdisplaybreaks
\usetikzlibrary{arrows.meta, positioning, calc}
\usepackage{amsmath}   % per aligned e altri ambienti
\usepackage{amssymb}   % per simboli extra come \mathbb, \dots
\usepackage{xcolor}

\newcommand{\hh}[1]{``#1''}
\renewcommand{\d}{\text{d}}
\AtBeginDocument{\RenewCommandCopy\qty\SI}

\usepackage{setspace}
\begin{document}

\title{Implementation and verification of a new free boundary module for Resistive Wall Modes simulations with the nonlinear 3D full MHD code SpeCyl}

\author{Francesco Galeazzi}
\email{francesco.galeazzi@igi.cnr.it}
\affiliation{Consorzio RFX, Padua, Italy}
\affiliation{Politecnico di Milano, Milan, Italy}

\author{Daniele Bonfiglio}
\affiliation{Consorzio RFX, Padua, Italy}
\affiliation{CNR-ISTP, Padua, Italy}

\author{Luca Spinicci}
\affiliation{UKAEA, Culham, UK}

\author{Susanna Cappello}
\affiliation{Consorzio RFX, Padua, Italy}
\affiliation{CNR-ISTP, Padua, Italy}

%\keywords{Resistive Wall Modes (RWMs), nonlinear 3D MHD, tokamak, thin resistive shells, verification}

\begin{abstract}

Self-consistent numerical simulations of free boundary instabilities, such as resistive wall modes (RWMs) and external kink modes, require special attention given their relevance on the path towards magnetic fusion based nuclear reactors. In this work, a new implementation and a verification benchmark are presented for the nonlinear three-dimensional (3D) full 
magneto-hydrodynamics (MHD) code SpeCyl, which self-consistently combines a free flow 3D treatment for the edge plasma velocity with double resistive wall thin shell magnetic boundary conditions, capable of capturing the essential features of common setups in experimental devices. Two coaxial resistive shells surround the plasma. Vacuum conditions are assumed between and outside the resistive shells, whose relative positions and time constants can be varied on purpose.

The implementation is extensively verified against Shafranov's analytic theory, valid for a circular cross section large aspect ratio tokamak and assuming a flat current profile. The correctness of newly implemented Boundary Conditions (BCs) is confirmed by the good agreement obtained for the linear growth rates of $(m=2,n=1)$ RWM and external kink, over a wide range of wall positions, resistive shell time constants, and for different initial reference equilibria. The $(m=2,n=1)$ RWM growth rates are also benchmarked against previous different RWMs modelling and successful validation studies for RFX-mod tokamak plasmas. The new BCs make SpeCyl a flexible and robust tool to investigate free boundary MHD modes in large aspect ratio magnetic confinement configurations, such as those achieved with the RFX-mod machine and expected in RFX-mod2.

\end{abstract}

\maketitle

\section{Introduction}

One of the main concerns for a thermonuclear fusion reactor is the uncontrolled growth of magneto-hydrodynamic instabilities, potentially capable of leading to a fast termination of the plasma discharge, with ensuing severe electromagnetic, thermal and mechanical loads dissipated on the reactor components in a short time. This is the case of tokamak plasmas, where major disruption events\cite{Bandyopadhyay2025MHD,TDS}, possibly triggered by free boundary instabilities\cite{Biskamp,Granetz1996Disruptions,Wesson,XTOR-2F} such as Vertical Displacement Events (VDEs), external kink modes and Resistive Wall Modes (RWMs), can terminate the plasma discharge and degrade the first wall of the confining machine \cite{Biskamp,Siccinio2022}. 
%Disruptions and sawtooth collapse in tokamaks are initiated by precursor phases dominated by specific classes of MHD instabilities: namely, the $(2,1)$ tearing mode and the $(1,1)$ kink mode. 

In this work, the focus is on the numerical implementation and verification analysis of linear growth rates for the specific instability class of RWMs, i.e., modes which would be stable in presence of a perfectly conducting boundary but become unstable on the characteristic resistive time scale of a realistic resistive boundary. These unstable modes play a key role in magnetic fusion machines\cite{Gimblett_1986, Chu2002RWM} and require stabilizing action from external control systems. Active feedback control systems\cite{walker2015tokamakcontrol} of external coils are currently intensively pursued\cite{Bandyopadhyay2025MHD} to reproduce the effects of a highly conducting wall, after the seminal proposal by Bishop\cite{Bishop1989IntelligentShell}, which was initially developed for a Reversed Field Pinch (RFP) configuration.

More specifically, this work aims at implementing and verifying a new boundary module in the nonlinear visco-resistive MHD numerical code SpeCyl\cite{Cap_Bisk_96}, motivated by the goal of increasing the realism of the electromagnetic boundary front-end to enable the investigation of MHD instabilities such as the RWMs\cite{Strauss2004ResistiveWall,Jones2026PIXIE3D} and the Resistive Wall Tearing Modes. The new setup consists of two resistive shells implemented in the computational tool for visco-resistive MHD plasma simulations. The inner shell mimics plasma facing components, while a second coaxial resistive thin shell plays the role of the main conducting structure and enables the simulation of RWMs\cite{Gimblett_1986,Hender89,Schn90,Saad_2022}.

This study constitutes a mandatory step towards the introduction of feedback system modelling interaction with 3D nonlinear MHD. The inclusion of conductive/resistive structures is needed to correctly capture the growth rates of RWMs. In fact, mirror currents induced in the resistive shell by the evolution of RWMs play an essential role in slowing down such instabilities\cite{Liu2008Analytical,Fitzpatrick1993_rwm,Haines13}, with respect to no-wall conditions, to the time scale of the resistive time of the shell. Reliable modelling of the amplitude of mirror currents and of their exponential decay as magnetic perturbations penetrate the shell is also key for fusion reactor integrated modelling, to predict 
the electromechanical force density distribution that may result in loads that the structures must withstand\cite{Sparago2025JOREK_STARWALL,Zakharov2012_disruptions,verif_JOREK_NIMROD_M3DC1,Isernia2019}.

The new boundary module is verified in two stages. First, a verification against analytical solutions\cite{Liu2008Analytical} is presented, on a class of flat current tokamak equilibria in a cylindrical plasma. The study of linear growth rates for the ($m=2,n=1$) RWM is discussed, by varying the outer shell resistivity in SpeCyl from an ideal wall case (external kink mode limit) to increasingly large wall resistivity. Then, a benchmark study with respect to previous realistic tokamak successful modelling\cite{Zanca2012Feedback} related to RFX-mod is discussed, and serves to anticipate a future more systematic validation of SpeCyl against experimental results.

The paper is organized as follows: Section~\ref{sec:section2} presents an overview of the SpeCyl code, including the set of MHD equations and the flow boundary conditions. Section~\ref{sec:section3} introduces the thin shell treatment and the set of newly implemented equations, which are detailed in Section~\ref{sec:section4} at the numerical level. The analytical model assumed as a reference for the first verification activity is presented in Section~\ref{sec:section5}. The setup adopted for simulations, together with their general features, is introduced in Section~\ref{sec:section6} and the numerical results are described in Section~\ref{sec:section7}. The second verification benchmark against previous successful realistic RFX-mod modelling is presented in Section~\ref{sec:section8}, while a summary and final outlook are provided in Section~\ref{sec:section9}. Three appendices are included, which deal with: the definition of double vacuum magnetic fields (Appendix~\ref{app:appB}), the theoretical framework to produce a perturbative stability analysis in the rectified tokamak (Appendix~\ref{app:appC}) and the elements for the derivation of the so-called Shafranov's flat current model (Appendix~\ref{app:appD}).

\section{The SpeCyl code}\label{sec:section2}

SpeCyl is a viscoresistive non-linear three-dimensional full MHD code\cite{Cap_Bisk_96}, which simulates the evolution of the magnetic field $\boldsymbol{B}$ and of the flow field $\boldsymbol{v}$ by means of a predictor-corrector scheme within a spectral approach\cite{Spinicci2023_MHD,Galeazzi2025_SpeCyl}. Cylindrical coordinates $[r,\vartheta,z]$ are adopted to write such fields, obeying the following set of equations\cite{Biskamp}:
\begin{equation} \label{eq:mom}   
\rho\Big(\dfrac{\partial\boldsymbol{v}}{\partial t} + \boldsymbol{v}\cdot\nabla\boldsymbol{v}\Big) = \boldsymbol{J}\times\boldsymbol{B}+\rho\nu\Delta\boldsymbol{v},
\end{equation}
\begin{equation} \label{eq:ind}   
   \dfrac{\partial\boldsymbol{B}}{\partial t} = - \nabla\times\boldsymbol{E},
\end{equation}
\begin{equation} \label{eq:ohm}   
   \boldsymbol{E}= \eta\boldsymbol{J} - \boldsymbol{v}\times\boldsymbol{B},
\end{equation}
\begin{equation} \label{eq:current}     
   \boldsymbol{J}= \ \nabla\times\boldsymbol{B},
\end{equation}
\begin{equation} 
\label{eq:divlessB}     
  \nabla\cdot\boldsymbol{B}=0.
\end{equation}
The pressure term $\nabla p$ is neglected in the momentum equation, and the plasma density profile $\rho$ is assumed to be constant ($\d\rho/\d t=0$).% thus avoiding solving the continuity equation (which would anyway require an ad-hoc diffusion term to roughly preserve the initial profile). 
$\ \boldsymbol{J}$ is the plasma current density, $\nu$ is the kinematic viscosity, $\eta$ is the plasma resistivity and $\boldsymbol{E}$ is the electric field. The transport coefficients ($\eta$ and $\nu=\mu/\rho$, with $\mu$ the dynamic viscosity) are assumed to be scalar fields. All terms are set in dimensionless form, using suitable reference quantities: $\rho=\hat{\rho}/\rho_0$; $\boldsymbol{B}=\hat{\boldsymbol{B}}/B_0$; $\boldsymbol{v}= \hat{\boldsymbol{v}}/v_A$; $t=\hat{t}/\tau_A$; $\boldsymbol{J}=\hat{\boldsymbol{J}}\mu_0a/B_0$; $\boldsymbol{E}=\hat{\boldsymbol{E}}/v_AB_0$. In these expressions, hatted terms are in physical units and $\rho_0=\hat{\rho}|_{r=0}$, $B_0=|\hat{\boldsymbol{B}}|_{r=0,t=0}$, while $v_A$ is the Alfvén velocity and $\tau_A$ is the Alfvén time. Their expressions are (being $\mu_0$ the vacuum permeability and $a$ the minor plasma radius) $v_A=B_0/\sqrt{\mu_0\rho_0}$ and $\tau_A = \sqrt{\rho_0\mu_0} a/B_0$. Equations~\eqref{eq:mom} -~\eqref{eq:divlessB} depend only on two dimensionless parameters, namely the Lundquist number $\mathit{S}= 1/\eta$ and the viscous Lundquist number $\mathit{M}=1/\nu$, which regulate the importance of diffusion for both equations\cite{Cap_Bisk_96}. The MHD model presented above is numerically solved in the computational domain constituted by a straight periodic cylinder, so that $0\leq r\leq a$, $0\leq\vartheta\leq2\pi$, and $0\leq z\leq 2\pi R_0$ (being $R_0$ the major plasma radius). Periodic boundary conditions are enforced at the two bases of the cylinder as in toroidal geometry. Consistently, the equations are Fourier-transformed in the $[\vartheta, z]$ coordinates, and finite differences are adopted along the radial coordinate $r$ to discretize differential operators. The non-linearities inherent in the simulated equations and the time advancing are dealt with using a semi-implicit treatment and a predictor-corrector scheme\cite{harned1985semi, Schn87, Spinicci2023Nonlinear}.

Given that Equations~\eqref{eq:mom} -~\eqref{eq:divlessB} are a set of partial differential equations, suitable boundary conditions for $\boldsymbol{B,v}$ must be specified for their solution. The new magnetic BCs are extensively presented in Sections~\ref{sec:section3},~\ref{sec:section4}, while the flow BCs are briefly outlined hereafter. The fully three-dimensional free normal flow velocity BCs implemented in SpeCyl have already proved essential in modelling free boundary instabilities with remarkable accuracy\cite{Spinicci2023Nonlinear,Cappello2024Impact}. Such flow BCs allow for a full 3D spectrum of velocity modes at the plasma edge, which couples with the magnetic field spectrum and increases the numerical complexity of the scheme. Discrete convolutions in the Fourier space manage the coupling of velocity and magnetic field harmonics. The flow BCs in SpeCyl are derived according to the treatment given in Ref.\cite{Delz08} for the case of a zero-$\beta$ paramagnetic pinch equilibrium but could be generalized to finite pressure models, as pointed out in Ref.\cite{Spinicci2023Nonlinear}. Based on the axisymmetric pinch velocity contribution, SpeCyl adopts a general expression for the velocity modes, which has the shape of a $\boldsymbol{E}\times\boldsymbol{B}$ flow. An edge velocity proportional to $\boldsymbol{E}\times\boldsymbol{B}$ is retrieved from Ohm's law at $r=a$, taking its cross product with $\boldsymbol{B}$, i.e.:
\begin{equation} \label{eq:ohm_curl_B}
(\boldsymbol{E}\times\boldsymbol{B})_a = \eta_a(\boldsymbol{J}\times\boldsymbol{B})_a - (\boldsymbol{v}\times\boldsymbol{B})_a \times \boldsymbol{B}_a.
\end{equation}
The orthogonal velocity is obtained by assuming $(\boldsymbol{E}\times\boldsymbol{B})_a \gg \eta_a (\boldsymbol{J}\times\boldsymbol{B})_a$, while for the parallel velocity a natural choice (and easily generalizable) is to set a null value, 
\begin{equation} \label{eq:pinch_vel_parallel}
\boldsymbol{v}_{||}(a)= 0\, .
\end{equation}
This assumption ($||$ denotes the direction parallel to the total magnetic field at the edge) is justified because the parallel velocity in MHD instability dynamics is typically negligible. The edge velocity then reads:
\begin{equation} \label{eq:pinch_vel}
\boldsymbol{v}(a)= \boldsymbol{v}_{\perp}(a) = \dfrac{(\boldsymbol{E}\times\boldsymbol{B})_a}{|\boldsymbol{B}_a|^2}\, .
\end{equation}
The ordering assumption made in writing Equation~\eqref{eq:pinch_vel} can be intended as a convenient closure for Equation~\eqref{eq:ohm_curl_B}, suitable for the boundary of a warm plasma. Neglecting the Ohmic term that accounts for the diffusive-resistive flux, i.e. $\eta\boldsymbol{J}\times\boldsymbol{B}\sim 0$, means considering negligible electromagnetic forces on the $\boldsymbol{E}\times\boldsymbol{B}$ flow spatial scales and thus acting as a force-free condition at the plasma boundary. Then, the magnetic energy flux $|\boldsymbol{B}_a|^2\boldsymbol{v}_{\perp}(a)$ due to external modes is put in direct relation with the Poynting vector $\boldsymbol{S}=\boldsymbol{E}\times\boldsymbol{B}$ on the left hand side of Equation~\eqref{eq:ohm_curl_B}, which results particularly effective in studying ideal MHD modes that are influenced by external conductors, e.g. the RWMs. In the limit of a small amplitude plasma displacement (i.e. when a linear stability analysis is performed), the edge velocity $\boldsymbol{v}_{\perp}(a)$ of Equation~\eqref{eq:pinch_vel} can be decomposed along $[r,\vartheta,z]$, retaining the three components which couple to the 3D edge magnetic field by means of discrete convolutions. 

The following decomposition is adopted throughout the work for the electric, magnetic and velocity fields, where $\hat{\boldsymbol{u}}_r$ is the unit vector aligned with the radial direction, while the subscript \hh{t} addresses the components of the field that are tangential to the wall, i.e. $t=[\vartheta,z]$: $\boldsymbol{E}=E_r\hat{\boldsymbol{u}}_r+\boldsymbol{E}_t$, $\boldsymbol{B}=B_r\hat{\boldsymbol{u}}_r+\boldsymbol{B}_t$,
$\boldsymbol{v}=v_r\hat{\boldsymbol{u}}_r+\boldsymbol{v}_t$.
%It is remarked that the edge velocity components to enforce fluid BCs are based on nonlinear expression which uses convolution sums themselves to treat products of real quantities in spectral space \cite{Spinicci2023Nonlinear}. These take advantage of the iterative structure of the SpeCyl code.

\section{Derivation of magnetic BCs}\label{sec:section3}

The electromagnetic boundary in SpeCyl employs a thin shell treatment\cite{Gimblett_1986}, adopted to model conductive/resistive elements enclosing the plasma. Introducing the generic shell position $r_w$, the thin shell modelling assumes that the radial component of $\boldsymbol{B}$ is continuous throughout the thin shell, i.e.,
\begin{equation} \label{eq:basic_cont_B} B_r^{m,n}\big|^{r_w^+}_{r_w^-}=0.
\end{equation}
The symbols $r_w^+,\ r_w^-$ identify an infinitesimally displaced radial position (positive and negative) with respect to $r_w$, and the vertical bar denotes the associated difference of evaluations. The integration of the Ampère-Maxwell law across the shell\cite{Galeazzi2025_SpeCyl} yields the expressions for the electromagnetic fields in correspondence with the resistive thin wall. For a generic radial position $r_w$, thin shell relations\cite{Gimblett_1986} are obtained as:
\begin{equation} \label{eq:evol_Br}   
 \dfrac{\partial B_{r,w}^{m,n}}{\partial t}  = \dfrac{1}{\tau_w} \Big[\dfrac{\partial}{\partial r} \Big(r B_{r,w}^{m,n}\Big)\Big|_{r_w^-}^{r_w^+}  \Big],
\end{equation}
\begin{equation} \label{eq:final_E}   
\boldsymbol{E}_{t,w}^{m,n} =\dfrac{r_w}{\tau_w} \hat{\boldsymbol{u}}_r\times\boldsymbol{B}_{t,w}^{m,n}\Big|_{r_w^-}^{r_w^+}.
\end{equation}
In Equations~\eqref{eq:basic_cont_B} -~\eqref{eq:final_E} the spectral nature appears through the superscripts $(m,n)$, which denote Fourier harmonics in the azimuthal and axial periodic coordinates, while the subscript \hh{t} denotes the components of the tangential field, as stated in the previous section. A proof of the identities in Equation~\eqref{eq:evol_Br} making use of Green's functions can be found in Ref.\cite{Pustovitov_2008}. The physical properties of the thin wall are contained in the adjustable wall time constant $\tau_w$:
\begin{equation}\label{eq:walltimecnst}
  \tau_w= \mu_0r_w\delta_w\sigma_w.  
\end{equation}
In the previous expression, $r_w$ is the radial position of the shell, $\delta_w$ is its thickness, and $\sigma_w=1/\eta_w$ represents the conductivity of the shell. The thin shell treatment can be regarded as the limiting case of the thick-wall treatment \cite{Nalesso1980,Gimblett_1986,Hender89} when the thin shell ordering is valid, i.e.: $\delta_w\ll r_w \ll2\pi R_0$. Various experimental conditions of interest can be reproduced within a thin shell treatment depending on the time scales on which the $\boldsymbol{B}$ modes penetrate the solid conductors. These are determined according to the Alfvén time $\tau_A$, which sets the different regimes for the diffusion of magnetic field modes through solid walls, also introducing in the picture the target time for the simulation $t_{sim}$:
\begin{itemize}
  \item the condition $\tau_w \rightarrow+\infty$ defines the case of an infinitely conductive wall. In other words, $\tau_w\gg t_{sim}$ and thus the diffusion-less case of an ideal shell is addressed;
  \item the condition $\tau_A\ll\tau_w\ll t_{sim}$ reproduces the case of a wall having finite resistivity;
  \item the condition $\tau_w\ll \tau_A $ instead sets instant-diffusion features, thus implying a sort of vacuum-wall transparent to magnetic fields. %, able to mimic a free plasma-vacuum interface. 
  This condition, which was already used in Refs.\cite{Spinicci2023Nonlinear,Cappello2024Impact}, will be employed in this study (unless otherwise stated) for the magnetic boundary at $r=a$ to mimic a free interface between the plasma and the vacuum. %This is the condition mostly considered in  and also in the present study as long as the wall at $r_w=a$ is concerned, when only the second resistive shell (having a finite distance from the plasma boundary, i.e. $r_w=b$) is assumed to have a finite resistivity.
\end{itemize}

The description of the electromagnetic boundary is completed with the expression for the radial electric field, obtained by applying the divergence operator to Equation~\eqref{eq:final_E} and using the gauge invariance for $\boldsymbol{E}$:
\begin{equation} \label{eq:E_r_mn}   
 E_r^{ m,n} = \eta_w\dfrac{\tau_w}{r_w}\nabla\cdot\boldsymbol{E}_{t,w}^{m,n}\ .
\end{equation}
Equations~\eqref{eq:evol_Br},~\eqref{eq:final_E} naturally reproduce the field expressions for the limiting cases discussed above, namely the ideal wall and the transparent wall conditions. For an ideal shell, $\tau_w\rightarrow+\infty$ and so $B_{r,w}^{m,n}$ is fixed in time and $\boldsymbol{E}_{t,w}^{m,n}$ = 0. On the other hand, a transparent wall condition is characterized by $\tau_w\rightarrow0$ and, owing to the finiteness of other fields, the continuity of $\boldsymbol{B}_{t,w}^{m,n}$ and of the radial derivative of $B_{r,w}^{m,n}$ are implied.\\

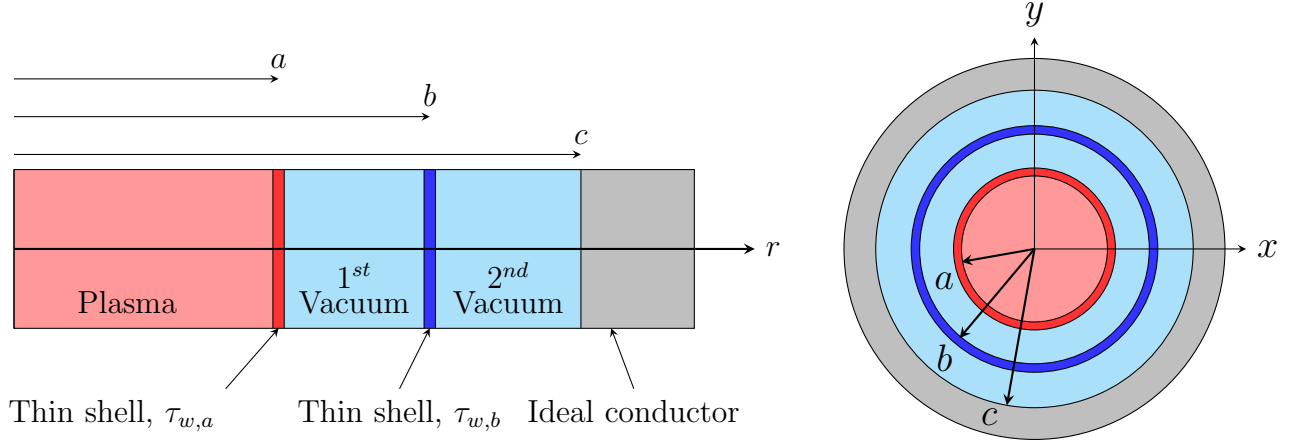
\begin{figure}[h!]
\centering
\begin{tikzpicture}[>=stealth, scale=1.0, transform shape]

% --- Parameters ---
\def\a{3.5}
\def\b{5.5}
\def\c{7.5}
\def\d{9.0}
\def\h{1.05}
\def\deltap{0.15}

% --- Region boundaries ---
\def\aL{\a-\deltap/2}
\def\aR{\a+\deltap/2}
\def\bL{\b-\deltap/2}
\def\bR{\b+\deltap/2}

% --- Setup label ---
\node[anchor=west,font=\bfseries] at (0,\h+1.8) {};

% --- Regions ---
\fill[red!40] (0,-\h) rectangle (\aL,\h);        % Plasma
\fill[red!80] (\aL,-\h) rectangle (\aR,\h);      % Thin shell 1
\fill[cyan!30] (\aR,-\h) rectangle (\bL,\h);     % Vacuum 1
\fill[blue!80] (\bL,-\h) rectangle (\bR,\h);     % Thin shell 2
\fill[cyan!30] (\bR,-\h) rectangle (\c,\h);      % Vacuum 2
\fill[gray!50] (\c,-\h) rectangle (\d,\h);       % Conductor

% --- Borders ---
\draw[thin] (0,-\h) rectangle (\d,\h);
\foreach \x in {0,\aL,\aR,\bL,\bR,\c}
  \draw[thin] (\x,-\h) -- (\x,\h);

% --- Axis ---
\draw[->,thick] (0,0) -- (\d+0.8,0);
\node[right] at (\d+0.8,0) {$r$};

% --- Labels ---
\node at (1.5,-0.7) {Plasma};
\node at ({(\aR+\bL)/2},-0.7) {Vacuum};
\node at ({(\bR+\c)/2},-0.7) {Vacuum};
\node at ({(\aR+\bL)/2},-0.7) [yshift=10pt] {$1^{st}$};
\node at ({(\bR+\c)/2},-0.7) [yshift=10pt] {$2^{nd}$};

% --- Distances ---
\draw[->] (0,\h+1.2) -- (\a,\h+1.2) node[left,above] {$a$};
\draw[->] (0,\h+0.7) -- (\b,\h+0.7) node[left,above] {$b$};
\draw[->] (0,\h+0.2) -- (\c,\h+0.2) node[left,above] {$c$};

% --- Annotations ---
\draw[->] (\a-0.7,-\h-0.8) -- (\a,-\h)
  node[pos=0, below left] {Thin shell, $\tau_{w,a}$};

\draw[->] (\b-0.4,-\h-0.8) -- (\b,-\h)
  node[pos=0,below] {Thin shell, $\tau_{w,b}$};

\draw[->] (\c+0.7,-\h-0.8) -- (\c+0.4,-\h)
  node[pos=0,below] {Ideal conductor};

% SETUP 2 — FRONT VIEW
\begin{scope}[
    xshift=13.5cm,
    scale=0.28,
    every node/.append style={transform shape=false,font=\large}
]

\node[font=\bfseries] at (0,4) {Front view};

\def\coreScale{0.7} \def\shellBoost{0.25}
\pgfmathsetmacro{\aLvis}{\aL*\coreScale}
\pgfmathsetmacro{\aRvis}{\aR+\shellBoost}
\pgfmathsetmacro{\bLvis}{\bL} \pgfmathsetmacro{\bRvis}{\bR+\shellBoost}

%\fill[red!40] (0,0) circle (\aLvis);
\fill[red!40] (0,0) circle (\aLvis);
\fill[red!80,even odd rule] (0,0) circle (\aRvis) (0,0) circle (\aLvis);
\fill[cyan!30,even odd rule] (0,0) circle (\bLvis) (0,0) circle (\aRvis);
\fill[blue!80,even odd rule] (0,0) circle (\bRvis) (0,0) circle (\bLvis);
\fill[cyan!30,even odd rule] (0,0) circle (\c) (0,0) circle (\bRvis);
\fill[gray!50,even odd rule] (0,0) circle (\d) (0,0) circle (\c);

\foreach \r in {\aLvis,\aRvis,\bLvis,\bRvis,\c,\d} \draw[thin] (0,0) circle (\r);

\draw[->] (0,0) -- (\d+1,0) node[right] {$x$};
\draw[->] (0,0) -- (0,\d+1) node[above] {$y$};

\foreach \radius/\label/\angle in {
  \a/a/190,
  \b/b/230,
  \c/c/260
} {
  \draw[thick,->] (0,0) -- ({1.0*\radius*cos(\angle)},{1.0*\radius*sin(\angle)});
  \node[anchor=north east] at ({0.95*\radius*cos(\angle)},{0.95*\radius*sin(\angle)}) {$\label$};
}
\end{scope} % front 2

\end{tikzpicture}
\caption{The SpeCyl double shell set-up, represented both with radial and axial views. The new implementation requires the modelling of a second resistive shell whose dynamics depends on $\tau_{w,b}$ and a second vacuum region, and their self-consistent coupling to other SpeCyl features.}
\label{fig:setups}
\end{figure}

Figure~\ref{fig:setups} illustrates the geometry of the new SpeCyl BCs: the plasma is surrounded by two coaxial cylindrical shells. The thin inner resistive shell (already present in Refs.\cite{Spinicci2023Nonlinear,Cappello2024Impact}) corresponds to the plasma edge at $r=a$ (this position is fixed by construction because the inner shell coincides with the outermost node of the plasma radial grid, as explained in Section~\ref{sec:section4}) and a second thin resistive shell is placed at $r=b$. The outer ideal shell is placed at $r=c$ with $c>b>a$ (typically far from the plasma) and serves to close the system of equations. Vacuum conditions are assumed in between the three coaxial conducting elements. The resistivity of the shells (as specified through the time constants $\tau_{w,a},\tau_{w,b}$) and their mutual distances are freely adjustable (by means of normalized positions $b/a,\ c/a$), allowing flexibility in investigating various experimental conditions. The presence of a second resistive shell of wall constant equal to $\tau_{w,b}$ requires not only to account for the evolution of the magnetic field components on it, but also to update the analytic expressions for magnetic field spectral components in the double coaxial vacuum. In addition, the double shell setup of Figure~\ref{fig:setups} must be self-consistently merged with the already existing 3D flow velocity treatment at the plasma edge, preserving a notable feature of SpeCyl\cite{Cappello2024Impact} with respect to other nonlinear MHD codes. A former set of BCs, implementing the hitherto described physical framework but only including the inner resistive shell in $r=a$ was non-linearly verified\cite{Spinicci2023Nonlinear} against the PIXIE3D code \cite{Chac08}. Together with a previous verification study \cite{PoP2010}, this proved the quantitative reliability of both SpeCyl and PIXIE3D codes over different test cases (tokamak and RFP configurations).

The new boundary module developed within this work is expected to increase the realism of the simulations performed with the new version of SpeCyl, especially when investigating RWMs. These instabilities require the presence of a resistive shell at finite distance from the plasma edge, where the free finite velocity couples with the magnetic field. The main equations arising from this approach are highlighted in the rest of this section, while in Appendix~\ref{app:appB} the analytical derivation of the double vacuum treatment is given in detail together with the development of the double vacuum coefficients. The starting point are the renewed expressions of the amplitudes of the components of $\boldsymbol{B}_{l}^{m,n}$, where $l=1,2$ identifies the vacuum region considered:
\begin{align}
B_{r,l}^{m,n} &= \mathcal{A}_l^{m,n}I'_{m}\Big(\dfrac{|n|r}{R_0}\Big) 
+ \mathcal{B}_l^{m,n}K'_{m}\Big(\dfrac{|n|r}{R_0}\Big),
\label{eq:Br} \\
B_{\vartheta,l}^{m,n} &= -\dfrac{mR_0}{|n|r}\Big[ 
\mathcal{A}_l^{m,n}I_{m}\Big(\dfrac{|n|r}{R_0}\Big) 
+ \mathcal{B}_l^{m,n}K_{m}\Big(\dfrac{|n|r}{R_0}\Big)\Big],
\label{eq:Btheta} \\
B_{z,l}^{m,n} &= \dfrac{n}{|n|}\Big[ 
\mathcal{A}_l^{m,n}I_{m}\Big(\dfrac{|n|r}{R_0}\Big) 
+ \mathcal{B}_l^{m,n}K_{m}\Big(\dfrac{|n|r}{R_0}\Big)\Big].
\label{eq:Bz}
\end{align}
Equations~\eqref{eq:Br} -~\eqref{eq:Bz} provide the expressions for the vacuum magnetic field harmonics, understood to be in Fourier space, involved in the definition of the new BCs. These are now discussed for the general case $m,n\neq0$, and in Section~\ref{sec:section4} their discretized version will be presented. The form of BCs for the axisymmetric mode ($m=0,n=0$) remains unchanged and can be found in Ref.\cite{Spinicci2023Nonlinear}. The first two equations are a specific version of Equation~\eqref{eq:evol_Br} for the two thin shells in $r_w=a$ and $r_w=b$. The partial radial derivatives of $B_r^{m,n}$ are written as a combination of $B_r^{m,n}(a)$ and $B_r^{m,n}(b)$, resorting to the double vacuum coefficients $\mathcal{C}^{m,n}_{ljs}$ for the magnetic field harmonics. The index $l$ denotes the first or second vacuum region, while the indices $j,s$ are set as $a,b$, according to the procedure detailed in Appendix~\ref{app:appB}. In brief, for the second shell at $r_w = b$:
\begin{equation} \label{eq:matching_b_1}  \tau_{w,b}\dfrac{\partial B_{r}^{m,n}(b)}{\partial t}  = [\mathcal{C}_{2bb}^{m,n} - \mathcal{C}_{1bb}^{m,n}]\ B_r^{m,n}(b) - \mathcal{C}_{1ba}^{m,n}\ B_r^{m,n}(a),
\end{equation}
while for $r_w=a$ the form of BC results in:
\begin{equation} 
\label{eq:matching_a_1}   
 \begin{split}
 \tau_{w,a}\dfrac{\partial B_{r}^{m,n}(a)}{\partial t}  =&\  \mathcal{C}_{1aa}^{m,n}\ B_r^{m,n}(a) +  \mathcal{C}_{1ab}^{m,n}\ B_r^{m,n}(b) - \dfrac{\partial}{\partial r} \Bigl(r B_{r}^{m,n}(r)\Bigr)\Bigg|_{pl}\ .
\end{split}
\end{equation}
The subscript \hh{$\mathit{pl}$} addresses the edge of the plasma domain. The non-axisymmetric tangential magnetic field components are determined enforcing the continuity of the tangential electric field (along $\vartheta,z$), for each Fourier mode in $r_w=a$, using Equation~\eqref{eq:final_E}: 
\begin{equation} \label{eq:ohm_a_theta}\eta_a(\nabla\times\boldsymbol{B}^{m,n})_{\vartheta,a} - (\boldsymbol{v}\times\boldsymbol{B})_{\vartheta,a}^{m,n} = -\dfrac{a}{\tau_{w,a}}\Big(i Z_{1aa}^{m,n}B_r^{m,n}(a) 
+ i Z_{1ab}^{m,n}B_r^{m,n}(b) 
- B_z^{m,n}(a)\Big),
\end{equation}
\begin{equation} \label{eq:ohm_a_zeta}\eta_a(\nabla\times\boldsymbol{B}^{m,n})_{z,a} - (\boldsymbol{v}\times\boldsymbol{B})_{z,a}^{m,n} = \dfrac{a}{\tau_{w,a}}\Big(i\Theta_{1aa}^{m,n}B_r^{m,n}(a) 
+ i\Theta_{1ab}^{m,n}B_r^{m,n}(b) 
- B_\vartheta^{m,n}(a)\Big).
\end{equation}
In both of these expressions, the left hand side represents the tangential electric field on the plasma side derived from Ohm's law, while the right hand side is the thin shell expression, which employs double vacuum coefficients $\Theta^{m,n}_{1js}$ and $Z^{m,n}_{1js}$, detailed in the Appendix~\ref{app:appB}. The convolution product $\boldsymbol{v}\times\boldsymbol{B}$ over the entire spectrum of modes produces a nonlinear term that must be dealt with to obtain a linear system with a spectral approach, as discussed in Section~\ref{sec:section4}.

\section{Implementation of the new boundary module}\label{sec:section4}

The numeric implementation of BCs derived in Section~\ref{sec:section3} must comply with the nonlinear and spectral structure of SpeCyl. From now on, the superscript $k$ identifies the generic time-step of the iterative SpeCyl workflow\cite{Spinicci2023Nonlinear,Galeazzi2025_SpeCyl} and for each $k$, the newly implemented boundary module enforces the BCs by solving a system of $7$ equations in $7$ unknowns. Three discrete radial positions are involved in the formulation: two belong to the plasma radial mesh and are $r_{N-1}=a-\Delta r$ and $r_{N}=a $ (edge node), while the third is the location of the second resistive shell, i.e., $r_{N+1}=b$. In the previous expressions, $N$ is the total number of radial mesh nodes and $\Delta r$ is the uniform spacing of the plasma mesh nodes. Then, the  unknowns of the boundary module (for each mode and for each time step) are: 
\begin{equation}\label{eq:unknowns}
B_r(a-\Delta r),\ B_\vartheta(a-\Delta r),\ B_z(a-\Delta r),\ B_r(a),\  B_\vartheta(a),\  B_z(a),\ B_r(b).
\end{equation}
Three equations to determine them arise as a consequence of the backward step of the so-called Thomas algorithm\cite{quarteroni2010numerical}, a simplified form of Gaussian elimination that can be used to solve tri-diagonal systems of equations. In fact, the spectral approach combined with a finite differences scheme to discretize second order derivatives (Laplace operator) produces a tri-diagonal mode-dependent linear system. The Thomas algorithm is computationally less expensive than other methods\cite{quarteroni2010numerical}, although based on a forward and backward step. These steps set constraints in the way the mesh unknowns are processed, because the coefficients entering the linear system depend both on the radial position (mesh dependency) and on the mode indices $(m,n)$ (modal dependency). Three conditions expressed as linear relations linking the magnetic field unknowns at the edge mesh nodes $r_{N-1}=a-\Delta r$, $r_{N}=a$ arise. They are constraints of numerical nature that propagate the boundary values to the inner plasma column radial mesh points:
\begin{align}
B_r^{m,n,k+1}(a - \Delta r)
&=
a_1^{m,n}\, B_r^{m,n,k+1}(a)
+ b_1^{m,n}\, B_\vartheta^{m,n,k+1}(a)
+ c_1^{m,n},
\label{eq:Br_thomas} \\
B_\vartheta^{m,n,k+1}(a - \Delta r)
&=
a_2^{m,n}\, B_r^{m,n,k+1}(a)
+ b_2^{m,n}\, B_\vartheta^{m,n,k+1}(a)
+ c_2^{m,n},
\label{eq:Btheta_thomas} \\
B_z^{m,n,k+1}(a - \Delta r)
&=
a_3^{m,n}\, B_z^{m,n,k+1}(a)
+ c_3^{m,n}.
\label{eq:Bz_thomas}
\end{align}
The full expression of the modal coefficients is omitted here because they are already encoded in the previous versions of SpeCyl and have not been modified. 

The remaining four equations are obtained starting from the equations introduced in the last part of Section~\ref{sec:section3}. The Equations~\eqref{eq:matching_b_1},~\eqref{eq:matching_a_1} are discretized in time using a first order Euler scheme of implicit type, so that their main unknowns become the Fourier components of $B_r^{k+1} (b)$ and $B_r^{k+1} (a)$, respectively. Moreover, Equation~\eqref{eq:matching_a_1} contains a radial partial derivative which is discretized according to a first order finite difference written on the plasma mesh nodes. They result in: 
\begin{equation}\label{eq:first_disc}
\tau_{w,b} \frac{B_r^{m,n,k+1}(b)-B_r^{m,n,k}(b)}{\Delta t} 
= (\mathcal{C}_{2bb}^{m,n} - \mathcal{C}_{1bb}^{m,n}) B_r^{m,n,k+1}(b) 
- \mathcal{C}_{1ba}^{m,n} \, B_r^{m,n,k+1}(a),
\end{equation}
and
\begin{equation}\label{eq:second_disc}\begin{split}
\tau_{w,a} \frac{B_r^{m,n,k+1}(a)- B_r^{m,n,k}(a)}{\Delta t} 
=& \left(\mathcal{C}_{1aa}^{m,n}- \frac{a}{\Delta r}\right) B_r^{m,n,k+1}(a) \\
+\ &\mathcal{C}_{1ab}^{m,n}\, B_r^{m,n,k+1}(b) 
+ \left(\frac{a}{\Delta r} -1\right) B_r^{m,n,k+1}(a-\Delta r).
\end{split}
\end{equation}
The route for the last two equations is longer: several steps are devised to access the algebraic version of Equations~\eqref{eq:ohm_a_theta},~\eqref{eq:ohm_a_zeta} that are compatible with the spectral nature of SpeCyl. First, the real space representation of $\boldsymbol{v}\times\boldsymbol{B}$ is split in two terms, to isolate the axisymmetric velocity components which couple to the generic magnetic field harmonic:
\begin{equation} \label{eq:convolution_v_b}  
 (\boldsymbol{v}\times\boldsymbol{B})_a^{m,n} = \boldsymbol{v}^{0,0}_a\times\boldsymbol{B}^{ m,n}_a + \sum_{(p,q) \neq (m,n)}\boldsymbol{v}^{m-p,n-q}_a\times\boldsymbol{B}^{\ p,q}_a.
\end{equation}
It is then required to obtain a linear equation for each mode and for each time-frame when the cross product is inserted back into Equations~\eqref{eq:ohm_a_theta},~\eqref{eq:ohm_a_zeta}. Moving to the spectral space (application of the Fourier operator $\mathcal{F}$) means defining an algebraic modal representation based on time dependent matrices. The terms in Equation~\eqref{eq:convolution_v_b} are suitably represented as a row-by-column product between the square matrix $\underline{\underline{\boldsymbol{v}_a}}$ and the column vector $\underline{\boldsymbol{B}_a}$ in the discrete Fourier space, where the contracted indices $(p,q)$ run over all the $(m,n)$ couples allocated in the simulation. To show this, the first contribution to the azimuthal component (along $\vartheta$) of $\boldsymbol{v}\times\boldsymbol{B}$ is considered (a similar procedure holds for the second, i.e., $v_rB_z$):
\begin{equation} \label{eq:vcrossb_t_fourier_first}
\mathcal{F}(v_z B_r)^{m,n}_a = \sum_{p,q}\Big(v_z^{m-p,n-q}\cdot B_r^{p,q}\Big)_a 
\Longleftrightarrow \hspace{0.2cm}(v_z^{0,0}\underline{\underline{I}}+ \langle \underline{\underline{v_z}}\rangle)\cdot\underline{B_r}\ .
\end{equation}
Equation~\eqref{eq:vcrossb_t_fourier_first} requires knowledge of the full magnetic field spectrum at a specific time-frame, meaning that the nonlinearity is not yet removed but isolated in the non-axisymmetric terms 
\begin{equation} \label{eq:non_axisym}
\langle\underline{\underline{\boldsymbol{v}}} \rangle\cdot\boldsymbol{B} = \sum_{(p,q) \neq (m,n)}\boldsymbol{v}^{ m-p,n-q}\times\boldsymbol{B}^{\ p,q}\ .
\end{equation}
A deconvolution procedure\cite{Spinicci2023_MHD} is adopted to treat them. This preserves the SpeCyl modal-based solution (serial solution of each Fourier harmonic) and allows individual inversion of each equation for the $(m,n)$ mode when enforcing the boundary conditions.
%given SpeCyl operates in the spectral space with the unknown at each radial mesh-point and at every time step being the $(m,n)$ modal component of the magnetic and flow fields. 
%The axisymmetric velocity components are numerically treated in an implicit way (both velocity and magnetic field are evaluated at the current time-step $k+1$), while the non-axisymmetric ones are dealt semi-implicitly (the non-axisymmetric velocities are evaluated at the prior time-step $k$). 
Taking advantage of the iterative structure of SpeCyl (see Refs.\cite{Spinicci2023_MHD,Galeazzi2025_SpeCyl}), the terms which couple with the non-axisymmetric velocity part in the convolution are evaluated explicitly, using their value at the prior time-step. So, taking into account the predictor-corrector advancement in time, the spectral character of SpeCyl and the deconvolution procedure yield the following expression, derived from Equation~\eqref{eq:vcrossb_t_fourier_first}, for the first component of the azimuthal part of the $\boldsymbol{v}\times\boldsymbol{B}$ already mentioned:
\begin{equation} \label{eq:vcrossb_t_first_a_k}
\mathcal{F}(v_z B_r)_a^{m,n,k+1} = \Big(v_z^{0,0,k+1}\cdot\underline{B_r}^{k+1} + \langle \underline{\underline{v_z}}\rangle^{k+1}\cdot\underline{B_r}^{k} \Big)^{m,n}_a\ .
\end{equation}
Equation~\eqref{eq:vcrossb_t_first_a_k} constitutes a single term of the cross product entering Equation~\eqref{eq:ohm_a_theta}. Using the expressions of the magnetic fields according to the coefficients listed in the final part of the Appendix~\ref{app:appB}, the spectral, time and space discretized representations of the Equations~\eqref{eq:ohm_a_theta},~\eqref{eq:ohm_a_zeta} then read:

\begin{equation}\label{eq:third_disc}
\begin{split}
\eta_a \Big(\frac{B_z^{m,n,k+1}(a)-B_z^{m,n,k+1}(a-\Delta r)}{\Delta r} 
- \frac{i n}{R_0}B_r^{m,n,k+1}(a)\Big) \\
+ \big(v_z^{0,0,k+1}\!\cdot\!\underline{B}_r^{k+1} 
+ \langle \underline{\underline{v_z}}\rangle^{k+1}\!\cdot\!\underline{B}_r^{k}\big)^{m,n}_a \\
- \big(v_r^{0,0,k+1}\!\cdot\!\underline{B}_z^{k+1} 
+ \langle \underline{\underline{v_r}}\rangle^{k+1}\!\cdot\!\underline{B}_z^{k}\big)^{m,n}_a \\
= \frac{a}{\tau_{w,a}} \big(i Z_{1aa}^{m,n}B_r^{m,n,k+1}(a) 
+ i Z_{1ab}^{m,n}B_r^{m,n,k+1}(b) 
- B_z^{m,n,k+1}(a)\big),
\end{split}
\end{equation}
and
\begin{equation}\label{eq:fourth_disc}
\begin{split}
\frac{\eta_a}{a} \Big(\frac{aB_\vartheta^{m,n,k+1}(a) - (a-\Delta r)B_\vartheta^{m,n,k+1}(a-\Delta r)}{\Delta r} 
- i m B_r^{m,n,k+1}(a)\Big) \\
- \big(v_r^{0,0,k+1}\!\cdot\!\underline{B}_\vartheta^{k+1} 
+ \langle \underline{\underline{v_r}}\rangle^{k+1}\!\cdot\!\underline{B}_\vartheta^{k}\big)^{m,n}_a \\
+ \big(v_\vartheta^{0,0,k+1}\!\cdot\!\underline{B}_r^{k+1} 
+ \langle \underline{\underline{v_\vartheta}}\rangle^{k+1}\!\cdot\!\underline{B}_r^{k}\big)^{m,n}_a \\
= \frac{a}{\tau_{w,a}} \big(i\Theta_{1aa}^{m,n}B_r^{m,n,k+1}(a) 
+ i\Theta_{1ab}^{m,n}B_r^{m,n,k+1}(b) 
- B_\vartheta^{m,n,k+1}(a)\big)\ .
\end{split}
\end{equation}
Equations~\eqref{eq:Br_thomas} -~\eqref{eq:second_disc} and~\eqref{eq:third_disc},~\eqref{eq:fourth_disc} constitute the system of $7$ equations that form the new boundary module: its solution, subject to an evident coupling along the coordinates $r, \vartheta$, defines the edge values of the magnetic field components, which are used by the code as BCs to solve the MHD model in the plasma domain. These values reflect the presence of the double shell/vacuum setup depicted in Figure~\ref{fig:setups}. Apart from the values of plasma quantities, updated at each time-step $k$, the hitherto described boundary module depends on constant parameters (such as the $\mathcal{C}^{m,n}_{ljs}$, $\Theta^{m,n}_{ljs}$ and $Z^{m,n}_{ljs}$ coefficients) that can be computed before the simulation starts. The only additional time-dependent spectral unknown is the radial magnetic field $B_{r}^{m,n,k}$ at the outer thin shell in $r_{N+1}=b$. This additional dynamic variable is employed for the vacuum field calculations (see Appendix~\ref{app:appB}) at each time-step, and complies with the
spectral nonlinear implementation of the SpeCyl.

\section{The Shafranov's flat current model (FCM)}\label{sec:section5}

The Shafranov's flat current model \cite{Shafranov1970,Liu2008Analytical} (FCM) provides analytical solutions for the dynamics of kink modes and RWMs on simplified equilibria in a current-carrying pinch. It has already been used for numerical verification studies addressed by different nonlinear codes such as JOREK\cite{Hoelzl_2014,McAdams2014_Tokamak}, NIMROD\cite{NIMROD_BCs}, M3D-C1\cite{Ferraro2016MultiRegion}, XTOR-2F\cite{Marx2017} and SpeCyl\cite{Cappello2024Impact}.
%as a basis for the investigation of external kink modes and RWMs, arising in the presence of a resistive structure around the plasma edge. 
The model is based on the straight tokamak limit, achieved for a vanishing inverse aspect ratio $\epsilon=a/R_0\rightarrow0$, and it descends from the linearization of the ideal MHD equations around a given equilibrium, when assuming flat profiles for the plasma density $\rho$ and the plasma current $\boldsymbol{J}$. Details on the procedure are given in the Appendix~\ref{app:appC}. The analytical model allows an exact shape for the eigenfunction (or plasma displacement $\boldsymbol{\xi}$) to be obtained and produces the expression of a dispersion relation that yields the linear growth rate of the instabilities.

\begin{figure}[ht!]
    \centering
    \includegraphics[width=0.7\textwidth]{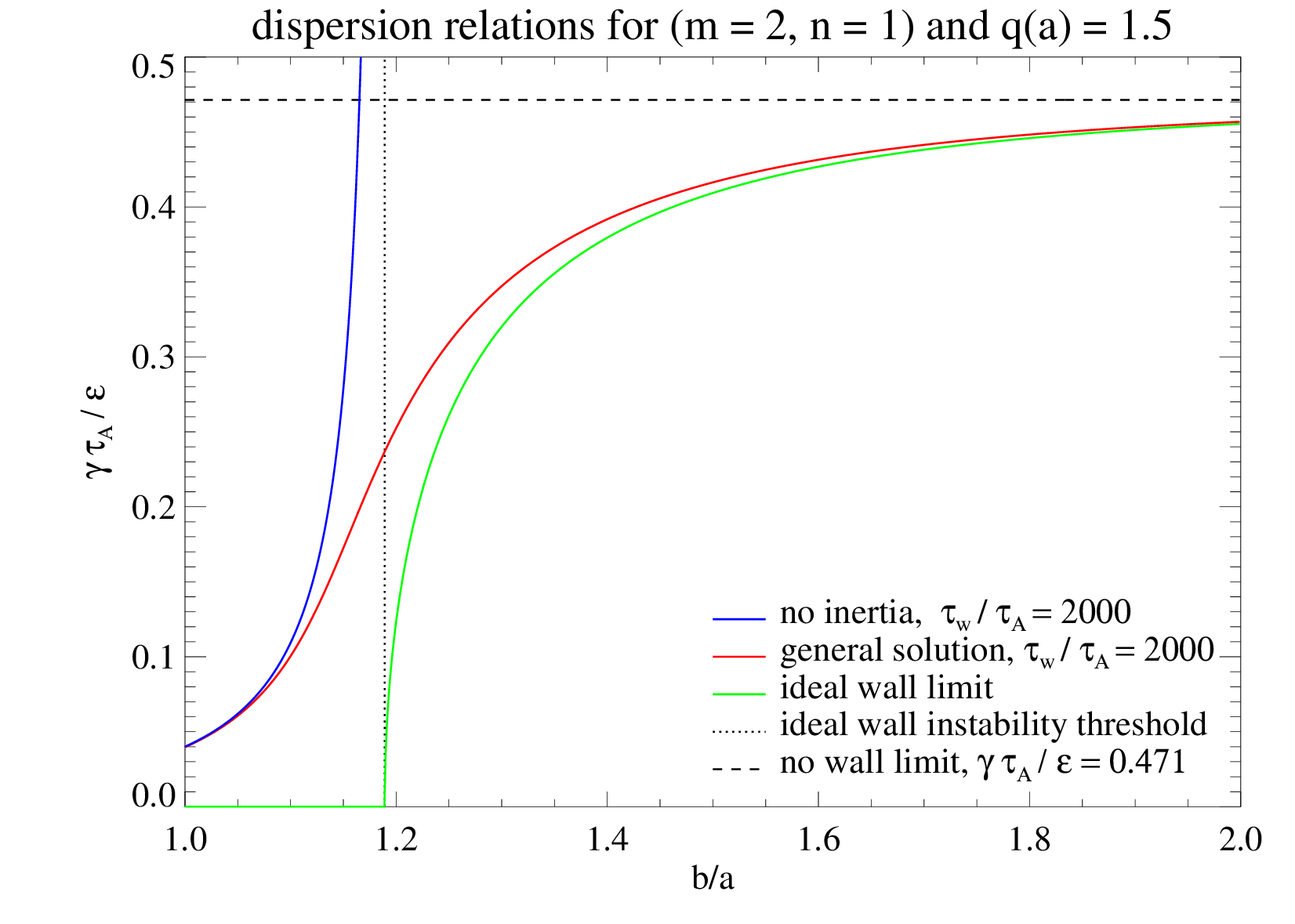}
    \caption{Dispersion relations from the linear analytical theory described in Section~\ref{sec:section5} for the mode $(m=2,n=1)$ and considering a large aspect ratio tokamak with flat current profile with $q(a)=q_0=1.5$. The dimensionless growth rate $\gamma\tau_A/\epsilon$, where $\epsilon$ is the inverse aspect ratio, is represented  as a function of the shell proximity $b/a$. The red curve is for a typical dimensionless wall time constant $\tau_w/\tau_A=2000$ and will be used for the benchmark with SpeCyl simulations. The green curve is obtained for the ideal wall limit when $\tau_w/\tau_A\rightarrow+\infty$. In this limit case, only the external kink instability is found, when the shell proximity exceeds the ideal wall instability threshold $(b/a)_{crit}=1.189$, marked by the dotted vertical line. Both the red and the green curves asymptotically converge to the so-called no-wall limit ($\gamma\tau_A/\epsilon=0.471$, horizontal dashed line) when the conducting shell is moved far away from the plasma. The blue curve is an approximation only valid when the plasma inertia can be neglected, i.e. for sufficiently low values of $\gamma$, when the wall position is close to the plasma edge.
    }
    \label{fig:gamma_tau}
\end{figure}

Although this model is already presented in the literature (e.g. in Ref.\cite{Liu2008Analytical}), its derivation is retraced here through its main steps, also enriched by the graphical outcomes of a purposely developed Newton algorithm to solve the general dispersion relation, as shown in Figure~\ref{fig:gamma_tau}. Two routes are available to derive the FCM: the first recasts the model into a particular version of the Energy Principle (a notable result of ideal MHD\cite{Wesson_1978,Freid}), while the second consists of solving a suitable Boundary Value Problem (BVP) for the so called magnetic flux function $\psi$, as done in Ref.\cite{Liu2008Analytical}. The two approaches can be proven equivalent because algebraic relations exist among their unknowns, as pointed out by Equations~\eqref{eq:link_1},~\eqref{eq:link_2}. The reason for the equivalence is the frozen-in-law of ideal MHD. Being the system periodic, the magnetic flux function, i.e. the linear perturbation to the equilibrium flux, is Fourier analysed in the $\vartheta,z$ coordinates, giving: 
\begin{equation}\label{eq:magfluxfnc}
\psi(r,\vartheta,z;t)= \Psi(r)\exp(im\vartheta - inz/R_0)\exp(\gamma t).
\end{equation}
This implies that the growth rate $\gamma$ will depend on the mode numbers $(m,n)$ of the considered MHD instability, in addition to relevant parameters such as the Alfvén time $\tau_A$ accounting for plasma inertia and wall time constant $\tau_w$. The equilibrium addressed to start the perturbative analysis enters the picture through the safety factor $q(r)$, defined as 
\begin{equation}\label{eq:safety_factor}
    q(r) = \dfrac{rB_z}{R_0B_{\vartheta}}.
\end{equation}
In the specific case of Shafranov's model, the safety factor turns out to be constant in the plasma domain, as a consequence of a uniform axial current density profile. All details are found in Appendix~\ref{app:appC}: departing from specific assumptions for the field profiles of interest, the growth rate dispersion relation is derived by solving a BVP for $\Psi(r)$ of Equation~\eqref{eq:magfluxfnc}.\\ 
The modal dispersion relation for MHD resistive wall modes and external kink modes is presented in Equation~\eqref{eq:rwm_full}, which yields the growth rate $\gamma$ of the MHD instability considered as a function of the uniform (over the radial coordinate in the plasma domain) q-value $q_0$, the resistive wall time constant $\tau_w$ and its relative position with respect to the plasma edge $r_w/a$\cite{Liu2008Analytical}:
\begin{equation}\label{eq:rwm_full}
   \dfrac{\nu}{(m - nq_0)} - \dfrac{1}{1 - \dfrac{\gamma\tau_w}{2\mu + \gamma\tau_w}\Big(\dfrac{a}{r_w}\Big)^{2\mu} } = \dfrac{1}{2}\dfrac{(\gamma \tau_A /\epsilon)^2q_0^2}{(m-nq_0)^2},
\end{equation}
where $\nu=sgn(m)$ and $\mu=|m|$. If no root of this equation exists with a positive real part, the plasma is stable. Compared to the corresponding expression in Ref.\cite{Liu2008Analytical}, Equation~\eqref{eq:rwm_full} presents a factor $1/\epsilon$ in the numerator of the right hand side, given the use of the \hh{poloidal} Alfvén time $\tau_A$. In addition, it differs by a factor $2$ in the fraction before the term $(a/r_w)^{2\mu}$, due to the adopted definition for $\tau_w$. Being Equation~\eqref{eq:rwm_full} a nonlinear expression of shape $f(\gamma,r_w/a,\tau_w )=0$, a Newton algorithm has been developed to compute the predicted growth rate $\gamma$, to compare it with the one obtained from the SpeCyl code with an equivalent setup. The general solution of Equation~\eqref{eq:rwm_full}, valid for both RWMs and external kink modes, is shown as a function of wall proximity $b/a$ in Figure~\ref{fig:gamma_tau} (red curve for RWM), for the case of the $(m=2,n=1)$ mode with $q(a)=q_0=1.5$.
%This curve generalizes the external kink mode accessed when an ideal wall is placed at a finite distance from the plasma edge. 
For an ideal wall, $\tau_w\rightarrow+\infty$ and so the corresponding dispersion relation reads
\begin{equation}\label{eq:ext_kink}
   \dfrac{\nu}{(m - nq_0)} - \dfrac{1}{1 - \Big(\dfrac{a}{r_w}\Big)^{2\mu} } = \dfrac{1}{2}\dfrac{(\gamma \tau_A /\epsilon)^2q_0^2}{(m-nq_0)^2}. 
\end{equation}
This is the specific case of the external kink mode that corresponds to the green curve in Figure~\ref{fig:gamma_tau}. This mode becomes unstable when the ideal shell is placed further from the plasma than the critical proximity $(b/a)_{crit}$, where $r_w=b$ is the position of the shell. 
The threshold just mentioned is obtained by imposing $\gamma=0, \tau_w\rightarrow+\infty$ in Equation~\eqref{eq:rwm_full} and reads
\begin{equation}\label{eq:threshold}
   (b/a)_{crit}=\Big(1 - \dfrac{m-nq_0}{\nu}\Big)^{-1/2\mu}. 
\end{equation}
For $q(a)=1.5$ and $(m=2, n=1)$, it follows that $(b/a)_{crit}=1.189$. Without any wall surrounding the unstable plasma, which means for $r_w\rightarrow+\infty$, the instability growth rate in the so-called no-wall limit is obtained. The latter obeys the expression indicated by the horizontal dashed line in Figure~\ref{fig:gamma_tau}:
\begin{equation}\label{eq:no_wall}
 (\gamma \tau_A /\epsilon)^2 = \dfrac{2}{q_0^2}(m-nq_0)[\nu - (m -nq_0)]. 
\end{equation}  
Finally, an approximation of Equation~\eqref{eq:rwm_full} is valid in the regime for which the growth rate is sufficiently small, i.e., when $\gamma\tau_A\ll1$ (which implies assuming the typical condition $\tau_A\ll\tau_w$ with $\gamma\tau_w\sim1$). This approximation neglects the role of plasma inertia: the condition coincides with having the MHD response of the system occurring on a timescale that results much slower than the Alfvén time. In such a case, the dispersion relation, shown by the blue curve in Figure~\ref{fig:gamma_tau}, becomes
\begin{equation}\label{eq:no_inertia}
 \gamma \tau_w  = -2\mu\dfrac{1-\nu(m -nq_0)}{\Big[1 -\nu(m-nq_0)-\Big(\dfrac{a}{r_w}\Big)^{2\mu}\Big]}. 
\end{equation} 
Only for small growth rates and when the wall position is placed close to the plasma edge, the agreement between the blue and the red curves is relatively good. In general, the plasma inertia cannot be neglected and plays an important role in the full dispersion relation of Equation~\eqref{eq:rwm_full}. However, Equation~\eqref{eq:no_inertia} is useful because it shows that in the RWM limit (i.e., when $\gamma\tau_A\ll1$) for a given mode, a given equilibrium, and wall position, the growth rate $\gamma$ is inversely proportional to the wall time constant $\tau_w$.

\section{Verification setup and preliminary tests}\label{sec:section6}

\subsection{Setup and consistency check}
\begin{table}[htbp]
  \centering
  \renewcommand{\arraystretch}{1.3}
  \setlength{\tabcolsep}{10pt}
  \begin{tabular}{|c|c|c|c|c|c|c|c|}
    \hline
    \rule{0pt}{4ex}
    $N_{\mathrm{modes}}$ & $\dfrac{m}{n}$ & $\epsilon=\dfrac{a}{R_0}$ & $\dfrac{\tau_{w,a}}{\tau_A}$ & $\dfrac{\Delta t}{\tau_A}$ & $N_r$ & $\mathit{M}$ & $\mathit{S}$ \\[1.5ex]
    \hline
    $5$ & $2$ & $0.05$ & $10^{-6}$ & $10^{-5}$ & $900$ & $10^{7}$ & $10^{7}$ \\
    \hline
  \end{tabular}
  \caption{Parameters adopted for the simulations with SpeCyl.}
  \label{tab:sim_param}
\end{table}
The numerical setup adopted in SpeCyl to reproduce the same conditions and profiles that characterize Shafranov's FCM described in the previous section is illustrated in Table~\ref{tab:sim_param}.

The inverse aspect ratio is set sufficiently small to retrieve the straight tokamak limit: $\epsilon=0.05$ is obtained by setting the plasma major and minor radius equal to $R_0=20$ and $a=1$, respectively. The resistivity $\eta$ and the kinematic viscosity $\nu$ of the plasma are assumed to be uniform along the radial coordinate. In addition, previous studies\cite{Cappello2024Impact} suggested setting both the plasma kinematic viscosity $\nu= 1/\mathit{M}$ and the plasma resistivity $\eta=1/\mathit{S}$ equal to $10^{-7}$, as the best compromise to approach ideal plasma conditions in a visco-resistive MHD code. The adopted time-step is $\Delta t/\tau_A=10^{-5}$ and the discretization along the radial coordinate employs $N_r=900$ mesh points equally spaced, selected after preliminary tests that showed satisfactory numerical convergence of the expected growth rate $\gamma$ for such a value. The numerical simulations consider a spectrum of harmonics composed by the equilibrium axisymmetric mode ($m=0,n=0$) and by $5$ non-axisymmetric components, characterized by a fixed helical pitch $m=2n$, for $1< n< 5$.

\begin{figure}[ht!]
    \centering
    \includegraphics[width=0.95\textwidth]{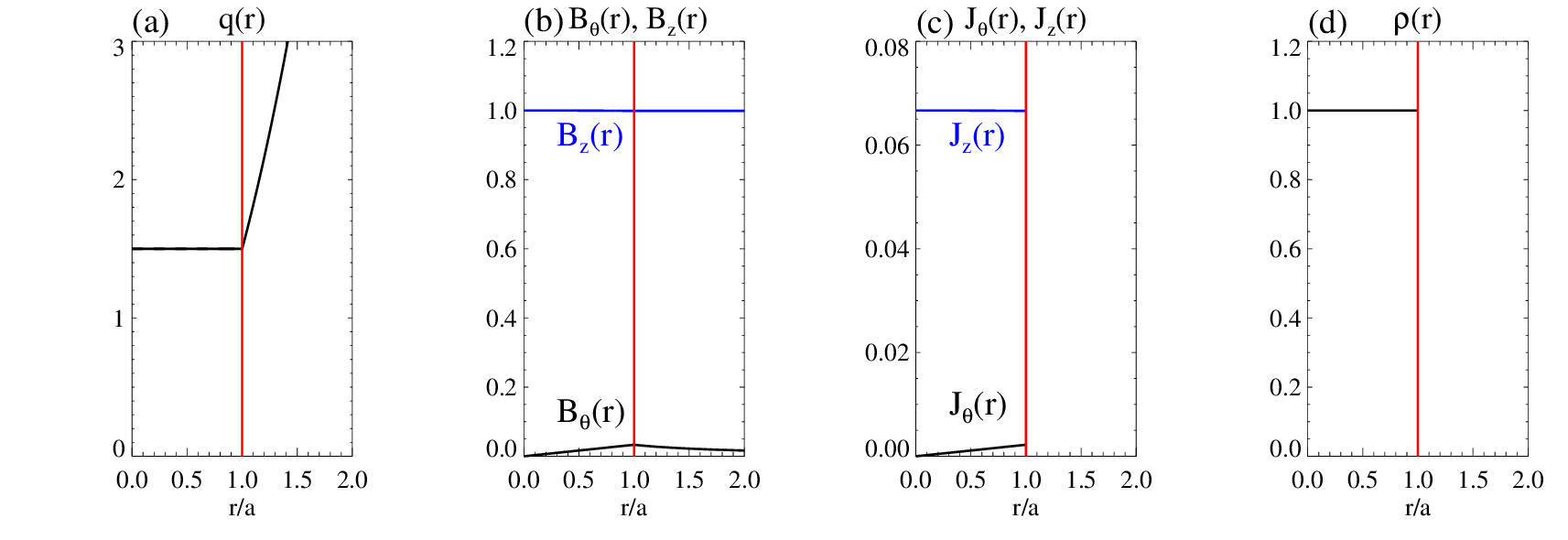}
    \caption{Equilibrium profiles for the verification of SpeCyl against the Shafranov's flat current model: panel (a) represents the circular cross section safety factor $q(r)$ according to Equation~\eqref{eq:sf_full}; the plasma domain $0\le r/a\le 1$ and the external vacuum region $r/a>1$ are separated by means of a red vertical line. The equilibrium magnetic fields are represented in panel (b), where the magnitude of the azimuthal field $B_\vartheta$ has the functional shape deduced in Equation~\eqref{eq:B_theta_full}. In panel (c), the plasma current density profiles are depicted. It is observed that $J_z=J_0\sim0.067$, which is consistent with the relation assumed in the derivation of the analytical model, i.e. $q_0=2B_0/\mu_0J_0R_0$, when $B_0=1$ and $\mu_0=1$ according to the SpeCyl normalization. Panel (d) represents the plasma density, assumed to be uniform in the computational domain.}
    \label{fig:eq_profiles}
\end{figure}

The equilibrium fields $\boldsymbol{B}, \boldsymbol{J}$ are self-consistently calculated in SpeCyl, combining Ohm's law and a force-free version of the momentum balance equation\cite{Delz08}. For ideal plasma conditions, the plasma current roughly scales with the (uniform) inverse plasma resistivity: Figure~\ref{fig:eq_profiles}(a-c) illustrates the resulting profiles obtained for $\boldsymbol{B},\boldsymbol{J}$ and $q$. Panel (d) also represents the adopted plasma density, which is assumed to be uniform over the plasma domain. A vertical red line marks the edge between the plasma region and the surrounding vacuum.

The analytical FCM assumes a free plasma-vacuum interface at $r=a$. To reproduce this condition in SpeCyl, the time-scale of the inner shell is set such that $\tau_{w,a}/\tau_A=10^{-6}$. This results in a transparent wall (instantaneous magnetic diffusion, negligible wall conductivity) consistent with the free boundary treatment, which has already been proven to be a suitable set-up for the modelling of free boundary MHD linear modes\cite{Cappello2024Impact}, due to the 3D velocity BCs (see Section~\ref{sec:section2}). Consequently, the second wall time constant $\tau_{w,b}$ (denoted as $\tau_w$ in the theoretical derivation of Section~\ref{sec:section5}) will be the only shell time constant whose values will be varied in the verification of the FCM discussed below, thus acting as the only effective conducting structure. 
%The modes included in the spectrum are then associated with the poloidal index $m=2n+\Delta m$, where $\Delta m\in [-2,2]$, for a fixed $n$: only five non-axisymmetric harmonics are included, corresponding to the $(m=2, n=1)$ mode and 4 helical harmonics up to the $(m=10, n=5)$.
%unstable eigenvalues obtained when combining the spectral basis harmonics generated by the rule presented before, for $n=1$.
In the rest of the paper, the attention is focused on the mode $(m=2,n=1)$. The initial equilibrium\cite{Delz08}, is destabilized by means of a velocity perturbation whose magnitude is set by $\mathcal{C}=10^{-6}$, which reads:
\begin{equation}\label{eq:perturbation}
  \delta v_r^{m,n}=\dfrac{\mathcal{C}}{r}\sin\Big(\dfrac{\pi r}a\Big)^{m+2}\cos(m\vartheta-nz/R_0). 
\end{equation} 
A consistency test is devised to check the effective numerical implementation of the set of equations that make up the new boundary module. Given a typical SpeCyl simulation of a RWM with the new boundary conditions developed in Section~\ref{sec:section4}, the test considers the continuity of $\boldsymbol{E}_t$ across neighbouring media for each time-step $\Delta t = t^{k+1}-t^k$ of the SpeCyl workflow: in correspondence with the plasma edge $r=a$, it compares the evolution of $E_\vartheta^{m,n},E_z^{m,n}$ when obtained according to the two routes onto which Equations~\eqref{eq:ohm_a_theta},~\eqref{eq:ohm_a_zeta} are built.
\begin{figure}[htp]
    \centering
    \includegraphics[width=0.6\textwidth]{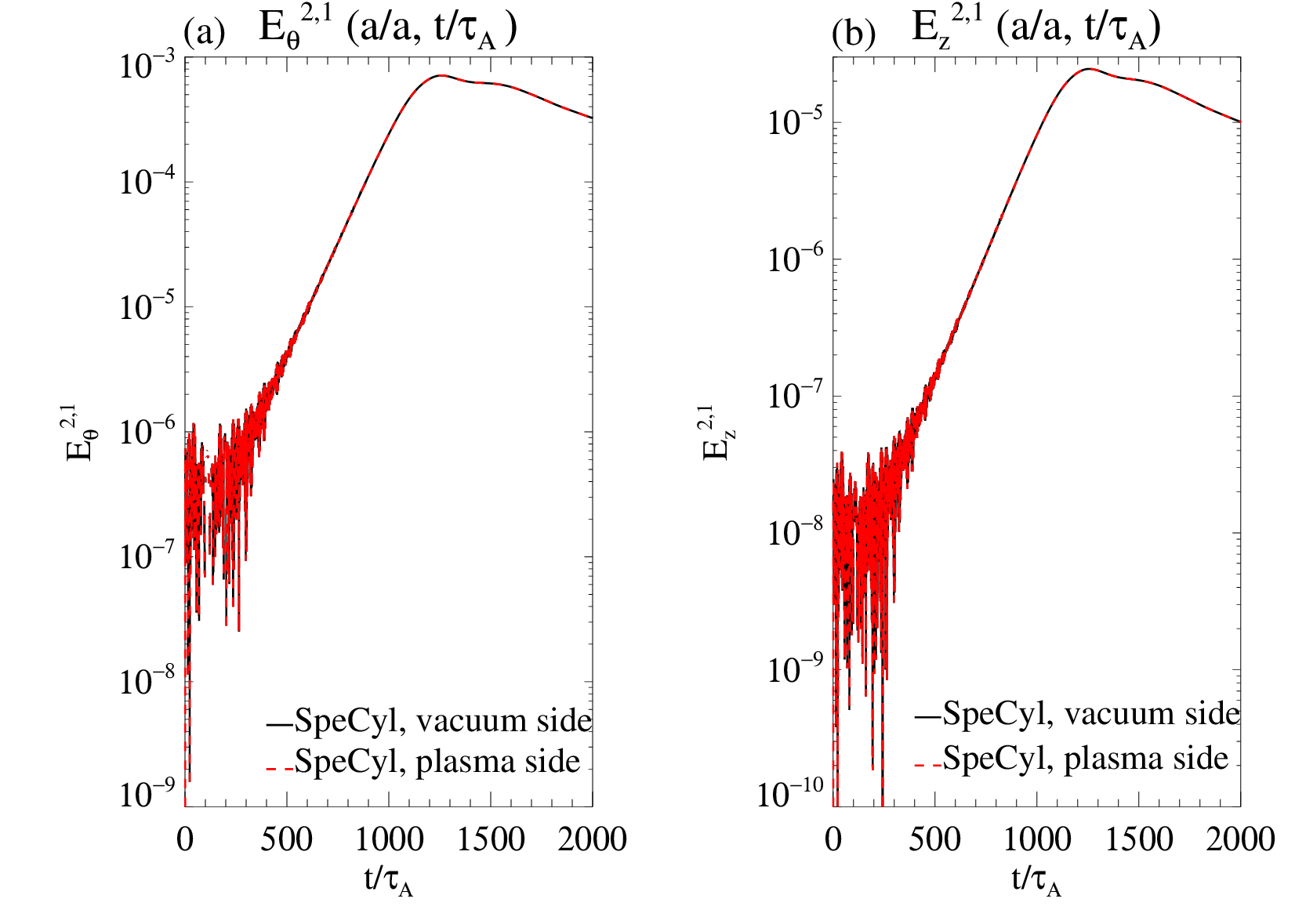}
    \caption{Comparison of the edge tangential electric fields $E_\vartheta^{m,n}$ in panel (a) and $E_z^{m,n}$ in panel (b), considering the case: mode $(m=2,n=1)$, safety factor $q(a)=q_0=1.5$, shell position $b/a=1.1$ and wall constant $\tau_{w,b}/\tau_A=1000$. For both the azimuthal and toroidal electric fields, the plasma expression in $r=a$ is compared to the vacuum expression in $r=a$, for the whole duration of the simulation.}
    \label{fig:compare_E}
\end{figure}
The plasma side expression (in dashed red) computes the tangential components via the generalized Ohm's law, reading:
\begin{equation}\label{eq:ohms_plasma}
\boldsymbol{E}_{t,a}^{m,n} = \eta_a(\nabla\times\boldsymbol{B}^{m,n})_{t,a} - (\boldsymbol{v}\times\boldsymbol{B})_{t,a}^{m,n},
\end{equation}
while the vacuum side expression (in black) results in a particular version of Equation~\eqref{eq:final_E}, which reads as follows:
\begin{equation}\label{eq:ohms_vacuum}
\boldsymbol{E}_{t,a}^{m,n} =\dfrac{a}{\tau_{w,a}}\hat{\boldsymbol{u}}_r\times\boldsymbol{B}_{t,a}^{m,n}\Big|^{a^+}_{a^-}.
\end{equation}
Equation~\eqref{eq:ohms_plasma} evaluates plasma fields at the time frame $t^{k+1}$, numerically determined according to the MHD equations solved in the plasma domain. Equation~\eqref{eq:ohms_vacuum} considers numerically computed plasma fields together with the resistive shell fields at $t^{k+1}$, and magnetic field expressions based on double vacuum coefficients $\Theta_{ljs}^{m,n}, Z_{ljs}^{m,n}$ presented in the Appendix~\ref{app:appB}. The test is particularly meaningful due to the intrinsic non-linearities involved in the time-stepping scheme\cite{Spinicci2023Nonlinear,Galeazzi2025_SpeCyl} and because the electric field is not directly evolved by SpeCyl. Equation~\eqref{eq:ohms_plasma} and Equation~\eqref{eq:ohms_vacuum} are here computed in post-processing and combine the multiple dependences on all modes of $\boldsymbol{B},\boldsymbol{v}$. In this sense, the agreement displayed in Figure~\ref{fig:compare_E} for the entire duration of the simulation tested, including both the linear growth phase and the nonlinear saturation phase, is significative. $E_{\vartheta,a}^{m,n}$ and $E_{z,a}^{m,n}$ are evolving almost identically when comparing their plasma side value and their vacuum side value. This confirms the new BCs are correctly framed into the SpeCyl iterative and modal dependent structure, determining a coherent evolution of the simulated quantities. The smooth behaviour in the saturation phase highlights the absence of edge field errors when the amplitudes of the harmonic components entering the convolutions are of comparable magnitude.

\subsection{Comparison of SpeCyl profiles with the linear theory}

\begin{figure}[htp]
    \centering
    \includegraphics[width=0.6\textwidth]{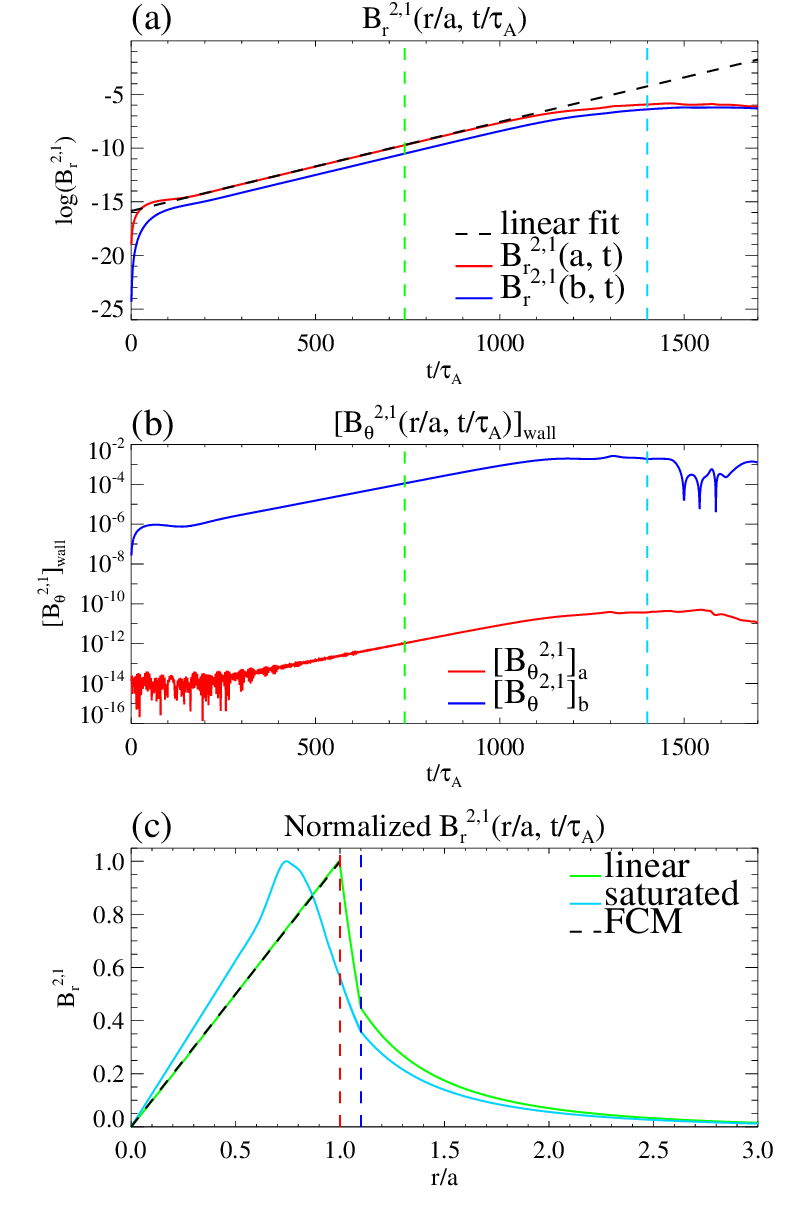}
    \caption{Features of a typical SpeCyl simulation using the new BCs, with $q(a)=1.5$, $\tau_{w,b}/\tau_A=1000$ and $b/a=1.1$. In panel (a), the time evolution of edge radial magnetic fields is shown. The red line denotes the $B_r^{2,1}$ at the $r=a$ radial position, while the blue line denotes the $B_r^{2,1}$ at $r=b$. A linear and a saturated stage are present in their evolution. The black dashed line represents the linear fit employed to estimate growth rates linear trend. The vertical green dashed line identifies a time snapshot in the linear stage and the cyan one has the same role for the saturated stage. The same colours are employed in the subsequent panels. Panel (b) adopts the same approach of panel (a) to represent $[B_\vartheta^{2,1}]_{r_w}$. %This quantity has been selected due to its link with the axial electric field $E_{z}^{2,1}$ and current density $J_{z}^{2,1}$. 
    Panel (c) shows the $B_r^{2,1}$ profile over the radial coordinate, for the two time snapshots in the linear and nonlinear phases, respectively. The linear behaviour matches the Shafranov's flat current model analytical expectations, indicated by the dashed black line. The cyan saturated profile displays a smoother behaviour at the plasma edge. The vertical dashed lines represent the two thin shells at $r=a$ (red) and at $r=b$ (blue).}
    \label{fig:evol_t_r}
\end{figure}

This subsection presents the dynamics and field profiles of a SpeCyl simulation with the new BCs (covering both the linear and nonlinear phases) and comparison with Shafranov's FCM linear analytical theory. The same simulation already introduced in the previous subsection is considered with $b/a=1.1$, $\tau_{w,b}/\tau_A=1000$, resulting in an unstable $(m=2,n=1)$ RWM. The outcome of the simulation is reported in Figure~\ref{fig:evol_t_r}. Panel (a) of Figure~\ref{fig:evol_t_r} shows the evolution in time of the value of $B_r^{2,1}(r,t)$ at the plasma boundary $r=a$ (red curve) and at the resistive shell $r=b$ (blue curve): a linear stage and a nonlinear (or saturated) stage can be identified in both curves. 
%The linear fit adopted to estimate the growth rates is represented with a black dashed line.
%together with two vertical dashed lines. These are important for the interpretation of other panels:
%The green vertical dashed line refers to a time frame $t_{lin}/\tau_A\sim740$ corresponding to the linear stage of the mode, while the cyan vertical dashed line marks its saturated stage $t_{sat}/\tau_A\sim1400$
Panel (b) displays the evolution in time of the azimuthal magnetic field discontinuity $[B_{\vartheta}^{2,1}]_{r_{w}}$ across the two shells, obtained as \begin{equation}\label{disc_BT}
[B_\vartheta^{2,1}]_{r_{w}} = B_\vartheta^{2,1}\big|_{r_w^+} - B_\vartheta^{2,1}\big|_{r_w^-}\ .
\end{equation}
For the plasma edge, the notation $r_w^-$ refers to the field value at $r=a$ in the plasma domain, while the notation $r_w^+$ refers to the vacuum field value at $r=a$. For the shell in $r=b$, $r_w^-$ refers to the field value computed in the inner vacuum region, while $r_w^+$ refers to the field values in $r=b$ as computed in the outer vacuum region. The azimuthal magnetic field discontinuity $[B_{\vartheta}^{2,1}]_{r_{w}}$ is directly related to the axial electric field $E_{z,w}^{2,1}$ because of Equation~\eqref{eq:final_E} and Ohm's law for the resistive shell $E_{z,w}^{2,1}=\eta_{w}J_{z,w}^{2,1}$ gives the corresponding current density.

The red curve in panel (b) of Figure~\ref{fig:evol_t_r} represents the discontinuity of $B_{\vartheta}^{2,1}$ for the shell in $r_w/a=1$ having the wall time constant set as $\tau_{w,a}/\tau_A=10^{-6}$: its magnitude over time is small ($\sim10^{-13}$), meaning that the azimuthal magnetic field is nearly continuous when comparing its plasma value and its vacuum value. This confirms the transparent wall nature of the inner shell, for which the magnetic field diffusion is almost instantaneous with respect to the plasma mode timescales. The blue curve in the same panel refers to the shell in $r=b$, for which $\tau_{w,b}/\tau_A=10^{3}$: the magnitude of the time evolution of $[B_{\vartheta}^{2,1}]_b$ is larger due to the finite conductivity of the shell.

Panel (c) of Figure~\ref{fig:evol_t_r} shows the radial profile of the magnetic field component $B_r^{2,1}$ (the eigenfunction) for both the plasma and the double vacuum region. Outside the plasma domain, the $B_r^{2,1}$ profile from SpeCyl is extended using a combination of Bessel functions that arise as a solution of the cylindrical Laplace problem (see Appendix~\ref{app:appB}) for the double vacuum region associated with $1< r/a \leq 2$. In the plasma domain $0 \leq r/a \leq 1$, the green profile, which refers to the linear stage of the mode, increases as a linear function in $r$. Indeed, the FCM (introduced in Section~\ref{sec:section5}) is also reported for comparison as a black dashed line: the predicted eigenfunction (see Appendix~\ref{app:appD}) is characterized by a very simple analytical solution, which goes linearly with $r$ and matches the simulation results.  

The radial derivative of the linear $B_r^{2,1}$ profile (in green) is discontinuous across the thin shell in $r=b$ with wall time constant $\tau_{w,b}/\tau_A=10^3$, due to the presence of a transient localized current. In fact, according to Equation~\ref{eq:evol_Br}, the evolution of the radial magnetic field across the resistive shell is driven by the slope imbalance of $B_r^{2,1}$ in the two regions. A resistive thin shell will then damp the currents over the penetration time $\tau_w$. The linear $B_r^{2,1}$ profile (in green) also shows an apparent discontinuity in its radial derivative for $r=a$, which is not expected due to the transparent wall nature of the inner shell. A closer inspection of the linear $B_r^{2,1}$ profile confirms that the discontinuity of the radial derivative is localized just inside the edge of the plasma domain ($r=a$), where surface currents\cite{Biskamp}, induced by RWM instability, occur. In the large aspect ratio limit, the axial component of such a current perturbation is predominant. The perturbation $J_z^{2,1}$ is associated with a discontinuity in $B_\vartheta^{2,1}$ and, due to the solenoidal condition for $\boldsymbol{B}^{m,n}$, it also determines the discontinuity mentioned above for $\partial(rB_r^{2,1})/\partial r$. The presence of surface current perturbations in the linear phase of free-boundary instabilities, such as external kinks and RWMs, is expected from analytical theory\cite{Zakharov_current} and was already discussed in the SpeCyl external kink simulations\cite{Cappello2024Impact}.

Finally, to conclude, in panel (c) the radial profile associated with a saturated time snapshot is represented in cyan: the linear character in the plasma is lost and the profile is smoother across the resistive shell in $b/a=1.1$ (even if the discontinuity is not completely removed), because the mode has almost completely penetrated it. %Smoothing also occurs in correspondence with the first resistive shell at the plasma edge, i.e. for $r/a=1$. 
In other words, the cyan profile in the $1 <r/a \leq 2$ region resembles the one valid for a single vacuum, as if the outer shell was not present.
%evident and the sustained current won't result decayed. This is confirmed by looking back at the panel (c), where the amplitude of the discontinuity is actually larger a the saturated time-frame.
%smoothing already highlighted due to the completed mode penetration (at saturation, timescales are sufficiently long), as if the second shell and the second vacuum were absent. The vertical discontinuities of the green curve come as a result of the thin shell approximation. The expected analytical behaviour is an edge Dirac delta distribution, i.e. $\delta(r/a-1)$, which is fully retrieved in SpeCyl simulations. The presence of a localized current expresses the mode development: the edge current allows the mode to continuously match the vacuum behaviour, accounting for the vacuum constraints $\nabla\cdot\boldsymbol{B}=0, \quad \nabla\times\boldsymbol{B}=0$. This defines the angular point already discussed in the continuous shape of $B_r^{2,1}$, for its linear stage.

\section{Extended verification study: comparison of growth rates}\label{sec:section7}

The verification of the new BCs in the SpeCyl code focuses on the $(m=2,n=1)$ mode, addressing the main parameters that affect the RWM stability. In particular, three tests are proposed to verify the correctness of the implementation, with the objective of reproducing the FCM growth rates. The first is based on SpeCyl simulations for three discrete wall time constants $\tau_w/\tau_A$, while $b/a$ spans a range between $1$ and $2$; the second test considers two values of $b/a$ and  a wide range for the dimensionless variable $\tau_w/\tau_A$. Both tests are associated with a specific equilibrium, $q(a)$ being fixed. The third test aims, instead, to demonstrate the robustness of the new boundary module against variations in the equilibrium configuration, in a similar way with respect to the analysis found in Ref.\cite{tanaka1973stabilizing}. The FCM determines a uniform safety factor profile across the radial coordinate; therefore, the variations of $q(a)$ correspond to different flat current values and equilibria. These tests add to previous verification studies of the SpeCyl code\cite{Spinicci2023Nonlinear,Cappello2024Impact}, including for the first time the RWM stability.

\subsection{Verification with respect to the shell proximity}

The linear growth rates resulting from SpeCyl are obtained by fitting the amplitude of the radial magnetic field $B_r^{m,n}(a,t)$ during the linear phase, as exemplified in panel (a) of Figure~\ref{fig:evol_t_r}. 
\begin{figure}[htp]
     \centering
    \includegraphics[width=0.7\textwidth]{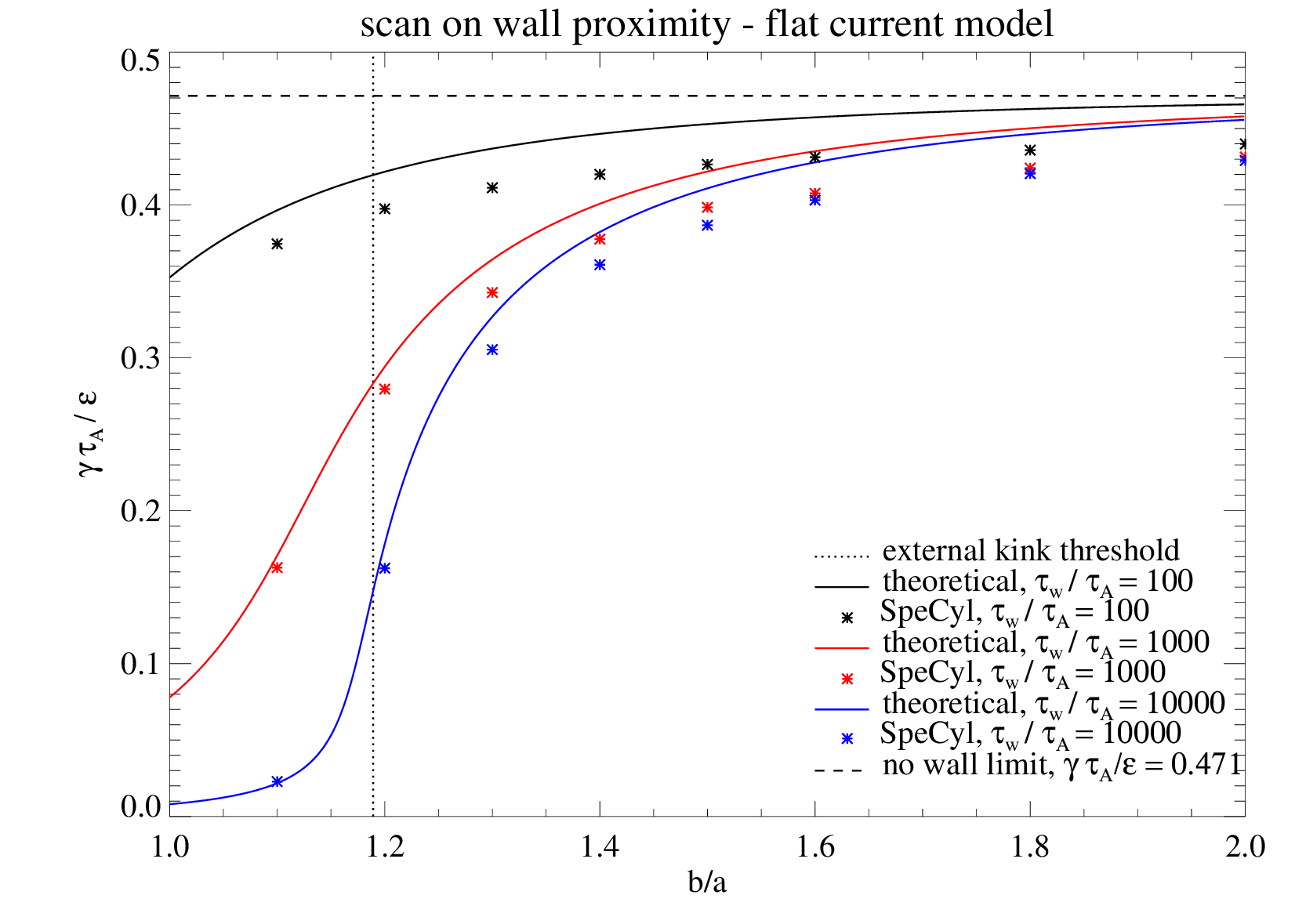}\caption{Comparison of growth rates estimates obtained by means of SpeCyl and the numerically solved dispersion relations.  The asterisks are the SpeCyl estimated growth rate values, while the three solid lines are the analytical predictions, obtained for a discrete $\tau_w/\tau_A$ each and in a range of $b/a$ between $1$ and $2$.
    The edge safety factor considered is $q(a)=1.5$ and the focus is on the $(m=2,n=1)$ mode. %The enforced inverse aspect ratio is $\epsilon=0.05$.
    }
    \label{figure:scan_in_boa}
\end{figure}
The results of the first test are shown in Figure~\ref{figure:scan_in_boa}: the new version of SpeCyl reproduces the analytical growth rates with good agreement, especially in the RWM regime when the wall position $b/a$ is close to the edge of the plasma. In addition, the threshold for the transition of the RWM into an external kink (around the critical proximity threshold $(b/a)_{lim}=1.189$, depicted as a black dotted vertical line in Figure~\ref{figure:scan_in_boa}) is correctly captured by SpeCyl growth rates: the solid blue line valid for $\tau_w/\tau_A=10000$ shows a steep increase in the growth rate value close to $r/a \geq1.189$, as expected from the change in the nature of the mode. The transition is progressively less steep for the other two curves with increasingly small wall time constants, because of the smaller difference in the growth rate between the RWM and the external kink mode regimes. For other cases, at larger $\gamma$ values and for smaller wall time constants (i.e. for low conductivity shells), the presence of an offset between the predicted growth rates and the simulated ones is observed. 

The relatively small offset reaches its asymptotic value for large $b/a$ values: this is evident when comparing the numerical and analytical no-wall limit growth rates, approached for increasing values of $b/a$. SpeCyl reproduces ideal MHD instabilities on slightly longer timescales with respect to the ideal plasma embedded in the analytic theory, especially in the external kink regime. This is consistent with the fact that the best agreement is achieved for relatively small growth rates, such as those of the RWMs.
%In other words, the absence of the inertial contribution provided by $\eta,\mu$ to the viscoresistive plasma is expected to play an active role for instabilities whose dynamics is faster and associated with \hh{large} growth rates.
A possible explanation for the observed offset of the numerical growth rates with respect to the analytical ones in the no-wall limit is related to the ideal plasma conditions ($\mu=0, \eta=0$) assumed by the analytical theory. The SpeCyl code solves a more complete set of equations that assume a visco-resistive plasma ($\eta$, $\mu$ are small but finite). It is therefore reasonable to expect the plasma non-idealities to introduce an additional effect with respect to the plasma inertia accounted for in the analytical theory. Indeed, as observed in Refs.\cite{Spinicci2023Nonlinear,Cappello2024Impact}, the agreement with ideal theory improves as the visco-resistive dissipation is reduced (i.e., when $\mathit{M,S}$ are increased), although the discrepancy could not be entirely removed.

\subsection{Verification with respect to the wall time constant}

\begin{figure}[htp]
    \centering
    \includegraphics[width=0.7\textwidth]{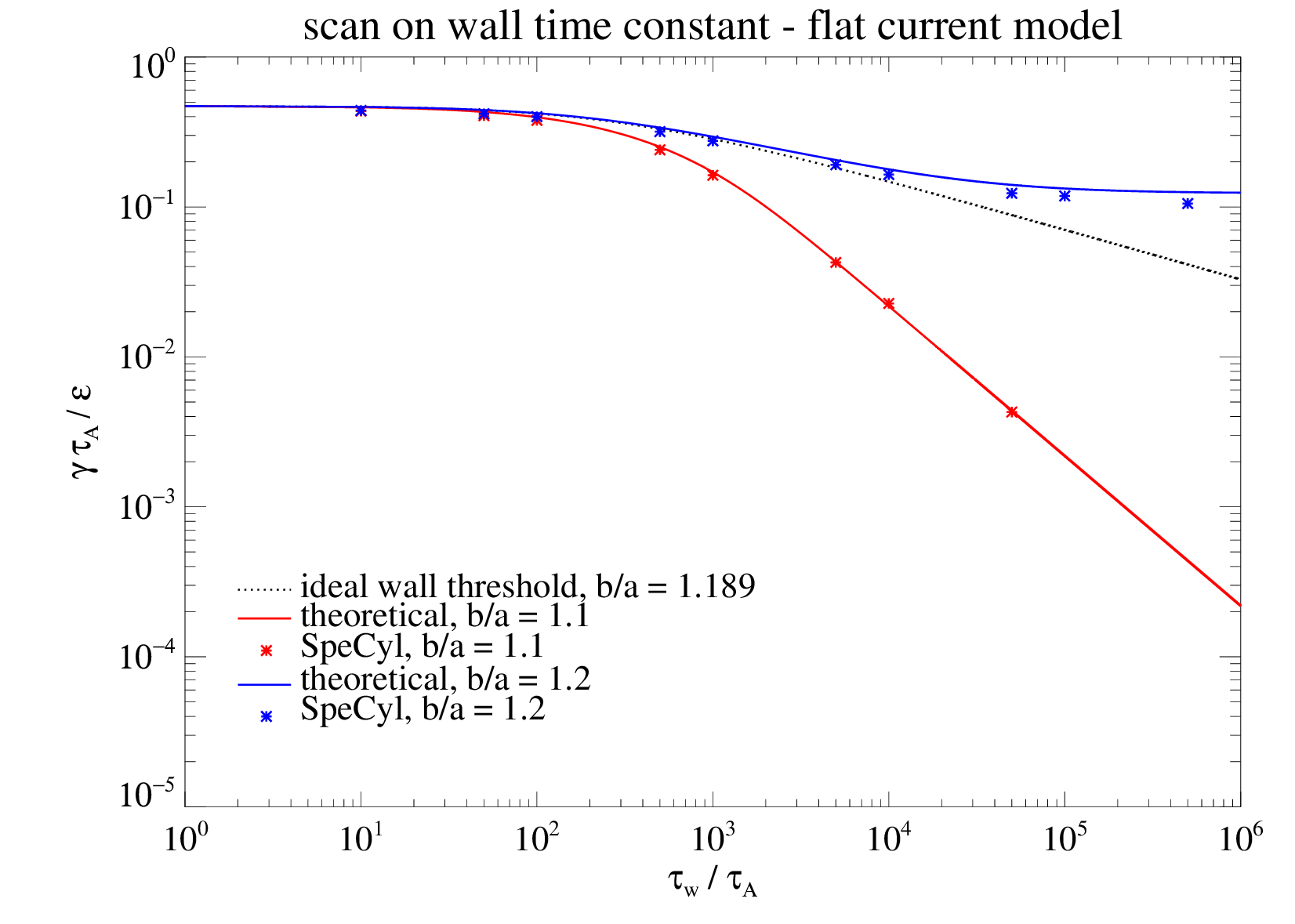}
    \caption{Comparison of growth rates estimates produce by SpeCyl and the numerically solved dispersion relations for two discrete values of $b/a$, with $\tau_w/\tau_A$ spanning over six decades. The solid lines are the analytic predictions, while the asterisks are the SpeCyl estimates. The edge safety factor is $q(a)=1.5$, for the mode $(m=2,n=1)$. 
    }
    \label{fig:scan_tau}
\end{figure}

The second test addresses the verification of the impact of the wall time constant. Figure~\ref{fig:scan_tau} confirms the good agreement shown by the first test. In this case, growth rates are reported for two discrete wall positions $b/a$ and with the wall time constant $\tau_w/\tau_A$ covering six decades. The bi-logarithmic representation is consistent with the one given in Refs.\cite{Hoelzl_2012,Ferraro2016MultiRegion} and allows the inverse dependence of $\gamma$ on the wall time constant $\tau_w$ to be shown when investigating a close-fitting configuration (see Equation~\eqref{eq:no_inertia}). Figure~\ref{fig:scan_tau} displays two solid curves: the blue corresponds to a wall position further from the plasma than the critical threshold $(b/a)_{crit}= 1.189$
(thus the $(m=2, n=1)$ kink mode still is unstable even if the wall was perfectly ideal), while the red is obtained for a wall position within such a threshold (thus the $(m=2, n=1)$ kink mode would be completely stable in the ideal wall limit: in this sense, only the red line represents a proper RWM). The critical threshold is also represented as a dotted black line. The growth rate estimates are in good agreement with the expected trend. %Simulations lasting up to $t_{sim}/\tau_A=10000$ were unable to highlight the unstable character of the mode, when looking at the  $b/a=1.1$ curve and in the large dimensionless wall time constant regime. The subtended time evolution of $B_r^{m,n}(1,t/\tau_A)$ confirms an oscillating behaviour, neither decaying nor growing. Such a feature is physically consistent: highly conductive walls have a stabilizing effect, at least on the time-scales investigated. Moreover, this last feature was anticipated by Figure~\ref{fig:gamma_tau}: the red curve for finite $\tau_w$ progressively collapses on the green curve as $\tau_w\rightarrow+\infty$, meaning the corresponding mode is stable (quasi-null growth rate) up to the proximity threshold.
%Taking a look at the subtended evolution in time, the observed onset of the unstable character is believed to overcome the oscillating nature of the mode at longer simulation times, departing from $t_{sim}=10000\tau_A$. The oscillating behaviour of such modes is the reason why two red asterisks are missing.
\begin{figure}[htp]
    \centering
    \includegraphics[width=0.7\textwidth]{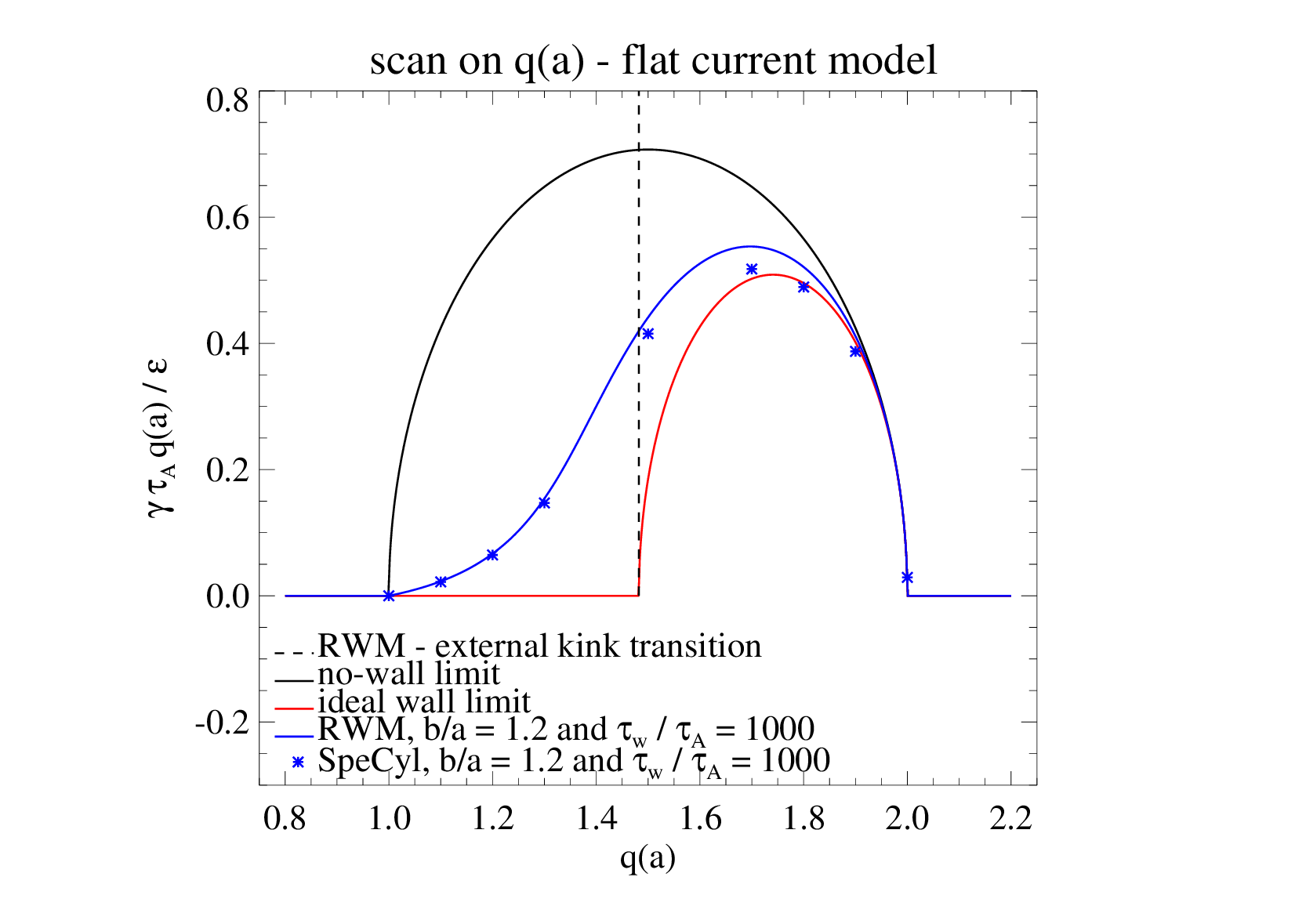}
    \caption{Comparison of growth rates estimates for the $(m=2,n=1)$ mode, obtained by means of SpeCyl (asterisks) and the numerically solved dispersion relation (blue curve) for $\tau_w/\tau_A =1000$ with $q(a)$ in the range between $1$ and $2$, for a fixed shell position $b/a=1.2$. The other solid lines are the limit cases analytical predictions: the black one corresponds to the no-wall limit, while the red one is the ideal wall limit. The black dashed vertical line represents the critical edge safety factor (uniform and equal to $q_0$ due to Shafranov's model assumptions), separating the equilibria destabilized by a RMW and those destabilized by an external kink.}  
    \label{fig:scan_qa}
\end{figure}

\subsection{Verification with respect to the reference equilibria}

The third verification test is reported in Figure~\ref{fig:scan_qa}. In this case, SpeCyl simulations are performed for varying $q(a)$, with fixed wall position $b/a=1.2$ and wall time constant $\tau_w/\tau_A=1000$. The quantity represented on the y-axis incorporates the value of the edge safety factor, according to the treatment given in Refs.\cite{Cappello2024Impact,tanaka1973stabilizing}. The new version of the code is confirmed to be capable of producing reliable growth rate estimates also as the initial equilibrium is varied. The average relative error is found around $6\%$, at most. In particular, a better agreement with the analytical growth rates is confirmed for the RWM regime (i.e., the range between $q(a)=1$ and $q(a)=1.48$). Figure~\ref{fig:scan_qa} represents a black dashed line for the edge safety factor value corresponding to the transition from RWM to an external kink: the threshold value is obtained by inverting the proximity threshold expression of Equation~\eqref{eq:threshold} and fixing the wall position $b/a$. For $b/a=1.2$, the discriminating edge safety factor is $q(a)_{crit}=1.48$: those equilibria characterized by $q(a)<1.48$ are associated with a RWM instability, while those equilibria for which $q(a)>1.48$ are prone to an external kink instability. 
In conclusion, the general result of the three tests is that the simulations performed with the new boundary module predict growth rates that agree well with those computed by the linear analytical model. In particular, the mean relative error is found around $6\%$, when analysing the entire set of simulations. A much smaller error around $3-4\%$ was found in the RWM regime with small growth rates. It should be noted that a similar magnitude of the discrepancy ($\sim 6\%$) was observed when investigating the results of simulations performed with the previous version of the code\cite{Cappello2024Impact}. In that case, using the external kink dispersion relation (Equation~\eqref{eq:ext_kink}) as a reference and considering $q(a)=1.5$, the maximum discrepancy was found around $6\%$, as in the present study. 
%Therefore, it can be concluded that the inclusion in the code of additional physics (given by the presence of the second resistive shell) is not responsible for a degradation of the agreement with the linear analytical theory. 
%In addition, SpeCyl does not account for the continuity equation: it describes, in principle, compressible plasma dynamics. This clashes with the derivation of the analytic dispersion relation, which indeed takes advantage of incompressibility to write the perturbed velocity $\delta\boldsymbol{v}$. Since conditions $\partial \rho/\partial t=0, \partial \rho/\partial r=0$ are enforced in SpeCyl simulations, the incompressibility constraint can be retrieved to hold at first order for $\delta \boldsymbol{v}$ when investigating the perturbed continuity equation. A posteriori verifications performed using the previous version of the code \cite{Spinicci2023_MHD}, confirmed $\nabla\cdot\boldsymbol{v}^{ 2,1}\approx 0$: even if the code equations avoid imposing incompressibility, this condition is respected in practice.This affects the fast instabilities in the Alfvénic growth rate regime. Such a feature is consistent with the observation reported in \cite{Spinicci2023Nonlinear,Cappello2024Impact}, where the numerical growth rate of the $(m=1,n=1)$ external kink in the no-wall limit was found to approach the analytical one for decreasing visco-resistive dissipation.

\section{Benchmark against RFX-mod device modelling}\label{sec:section8}

A second verification study of the new SpeCyl code is now presented, adopting a class of large aspect ratio equilibria initially studied by Wesson\cite{Wesson_1978,Wesson}, which replaces the FCM used for the just discussed extensive verification. The Wesson's model allows simulations with more realistic safety factor profiles and has been shown to reproduce (when realistic parameters for the conducting structures are employed) the spontaneous
$(m=2, n=1)$ RWM growth rates observed in the RFX-mod machine operated as a circular cross section low-current Ohmic tokamak\cite{Zanca2012Feedback}. The spirit of this section is to compare the SpeCyl linear growth rates prediction with those obtained with RFXlocking\cite{Zanca2009} simulations, using the same setup with current profile and shell parameters aiming at reproducing tokamak experiments in the RFX-mod device, as described in Ref.\cite{Zanca2012Feedback}. This activity constitutes a preliminary step in view of future validation studies\cite{GreenwaldPoP2010} of the new SpeCyl code developed within this work, directly against the RFX-mod experimental measurements.

\begin{figure}[htp]
    \centering
    \includegraphics[width=0.7\textwidth]{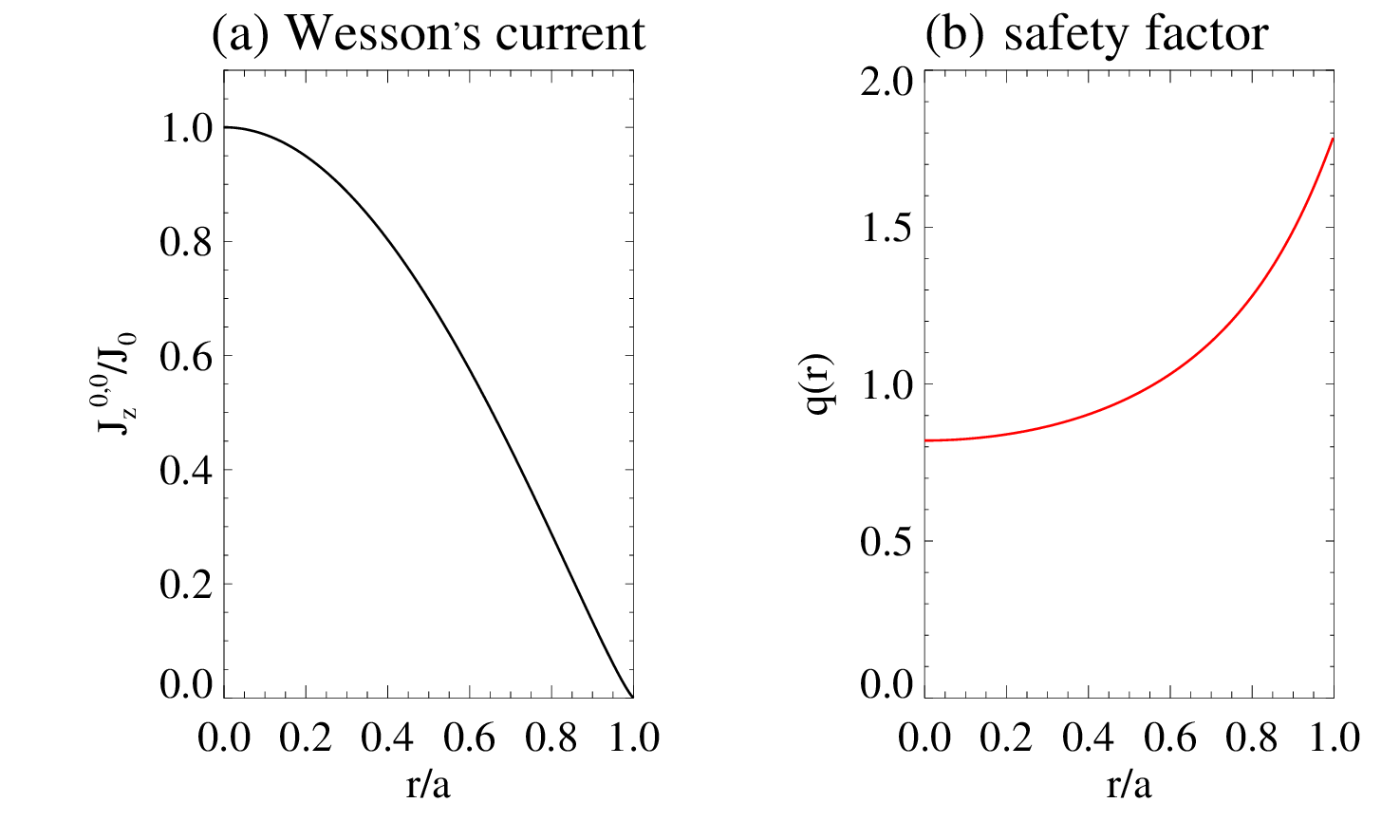}
    \caption{Wesson's profile for the equilibrium axial current $J_z^{0,0}$ and corresponding safety factor $q(r)$.}
    \label{fig:equil_W}
\end{figure}

RFXlocking is a cylindrical code for the solution of the MHD force-balance equation (Newcomb's equation) in the zero-pressure formulation, coupled with models for passive and active conductive structures surrounding the plasma. It was developed to model the feedback control of Tearing Modes (TMs) in the RFP configuration of the RFX-mod device\cite{Zanca2009} and later adapted to study both RWMs\cite{Zanca2012Feedback} and TMs\cite{Zanca2015} in RFX-mod tokamak discharges. The code is also used for other devices, such as JET\cite{Tonel2025}. In the study\cite{Zanca2012Feedback} targeted for this second verification test, the Wesson's current profile $J_z(r)^{0,0}=J_{z,0}(1 -(r/a)^2)^{\nu}$ was adopted with $\nu=1.2$, to obtain a safety factor profile consistent with estimates from target RFX-mod tokamak discharges, with $q(0)=0.82$ and $q(a)=1.78 $. These profiles are represented in Figure~\ref{fig:equil_W}. 
The code employed the realistic aspect ratio for RFX-mod (with plasma minor and major radius equal to $a=0.459\ \text{m}, R_0=2\ \text{m}$, respectively) and the actual position and time constants of the passive structures corresponding to the Inconel alloy vacuum vessel ($r_{w,a}= 0.490\ \text{m}$, $\tau_{w,a}=3\ \text{ms}$) and of the Passive Stabilizing Shell (PSS)\cite{Piovesan2013} made of copper ($r_{w,b}=0.5127\ \text{m}$, $\tau_{w,b}=100\ \text{ms}$). In the absence of feedback control, the natural growth rate of the $(m=2,n=1)$ RWM was calculated by RFXlocking to be equal to $\gamma=47\ \text{s}^{-1}$, very close to the value of $\gamma=44\ \text{s}^{-1}$ measured in the target RFX-mod tokamak discharge.

The same numerical setup has been reproduced in the SpeCyl code, with the main difference, with respect to RFXlocking, that the inner resistive wall is by construction located at the plasma radius $r=a$. For this reason, a scan in $\tau_{w,a}$ has been performed, from the most conservative assumption of a transparent wall ($\tau_{w,a}\ll\tau_A$), for which higher growth rates than in RFXlocking are expected and thus requiring a tighter control law, to the least conservative condition when the actual time constant of $3\ \text{ms}$ is used, as in RFXlocking, although with the wall touching the plasma. Regarding the outer conducting shell, the proper shell proximity $b/a=1.117$ is used, with three values of the PSS time constant: the nominal value of $100\ \text{ms}$ for a thin copper shell, the effective value of $125\ \text{ms}$ as estimated in RFX-mod by penetration of the vertical field ($m=1,n=0$) in vacuum (and interpreted as the effect of the outer mechanical stainless steel support structure\cite{Marchiori2007}), and an intermediate value of $115\ \text{ms}$. It is worth remarking that the shell time constants estimated in Ref.\cite{Marchiori2007} must be doubled for direct comparison with those of this work, as they are based on a definition that differs by a factor of $1/2$ from that of Equation~\eqref{eq:walltimecnst}. Such wall time constants are converted into dimensionless units, required in SpeCyl, by computing the Alfvén time $\tau_A$ consistent with the RFX-mod plasma parameters taken from Ref.\cite{Zanca2012Feedback} and listed in Table~\ref{tab:exp_quant}. The Alfvén time obtained is $\tau_A=1.97\cdot 10^{-7}\ \text{s}$. The resulting wall time constants in Alfvénic units are $\tau_{w,a}/\tau_A= 15051$ (corresponding to $3\ \text{ms}$) and  $\tau_{w,b}/\tau_A=501711$ (corresponding to $100\ \text{ms}$).
\begin{table}[h!]
    \centering
    \begin{tabular}{|c|c|c|}
     %\hline
      %Name & Symbol & Value & Units \\
     \hline
     \rule{0pt}{2.5ex}
     Core electron density & $n_e$&$2\cdot10^{19}\ \text{m}^{-3}$ \\
  Plasma current & $I_p$&$1.5\cdot10^5\ \text{A}$ \\
 Minor plasma radius & $a$&$ 0.459\ \text{m}$ \\
 Inverse aspect ratio  & $\epsilon$&$0.2295$ \\
 Edge safety factor & $q(a)$&$1.78$\\
    \hline
    \end{tabular}
    \caption{Experimental parameters for the target RFX-mod tokamak discharges for Resistive Wall Modes studies, as found in Ref.\cite{Zanca2012Feedback}.}
    \label{tab:exp_quant}
\end{table}
\begin{figure}[htp]
    \centering
    \includegraphics[width=0.4\textwidth]{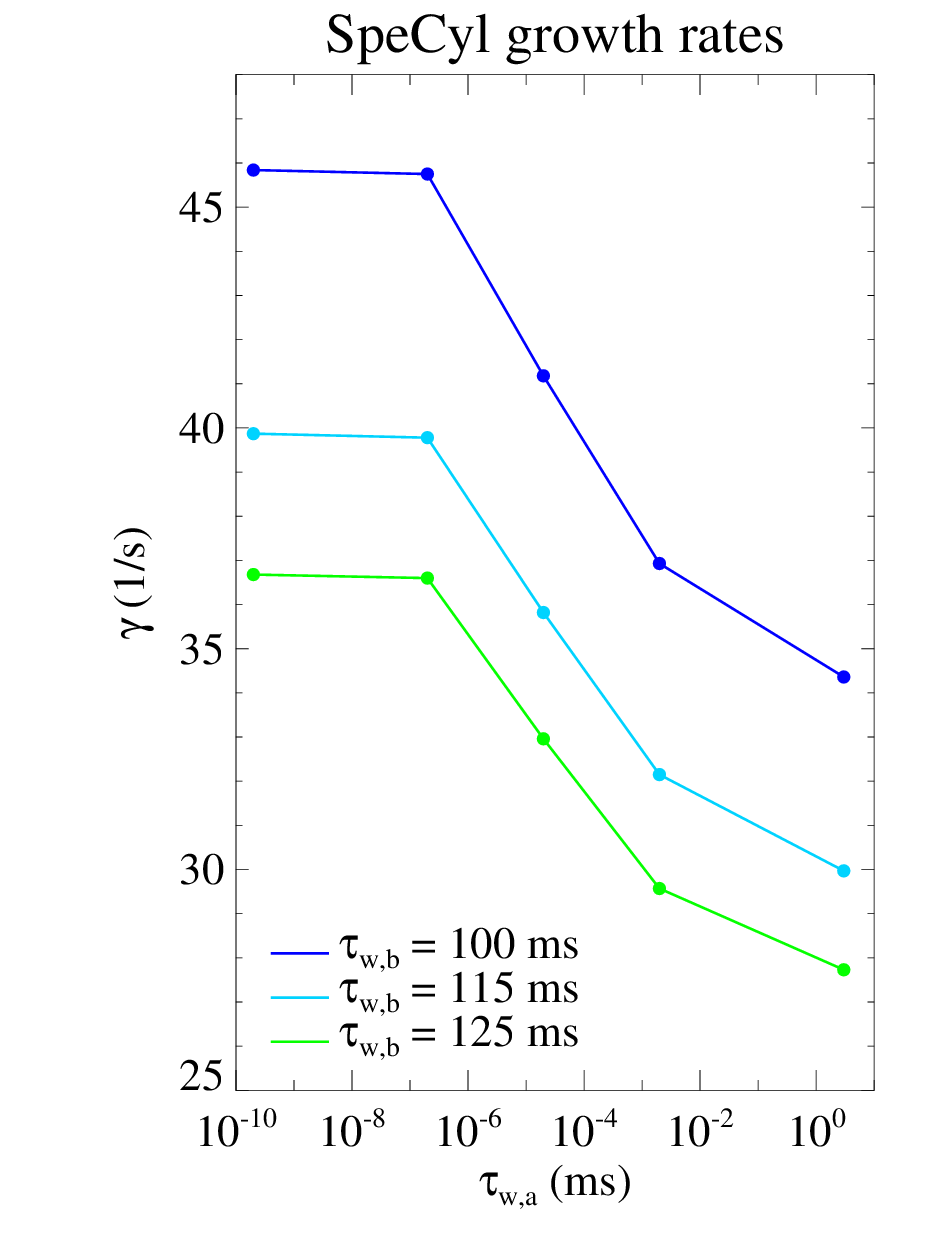}
    \caption{SpeCyl estimates for growth rates obtained with the setup in Ref.\cite{Zanca2012Feedback}.}
    \label{fig:est_rfxmod}
\end{figure}

The resulting growth rates, converted into physical units, are represented in Figure~\ref{fig:est_rfxmod}. The flexible setup characterizing the new version of SpeCyl allows for a sensitivity analysis. Looking from left to right, for transparent wall conditions at $r=a$ ($\tau_{w,a} \lesssim 10^{-7}\ \text{ms}$), $\gamma$ is insensitive to the inner shell and depends only on $\tau_{w,b}$, while, as $\tau_{w,a}$ grows to large values, $\gamma$ decays. %The three curves get closer with increasing $\tau_{w,a}$, converging to stability for $\tau_{w,b}\rightarrow\infty$.
The blue curve in Figure~\ref{fig:est_rfxmod} is the reference and obtained for a PSS with a wall time constant $\tau_{w,b}= 100\ \text{ms}$. The curve shows the growth rate behaviour as the first wall at the plasma edge moves from a transparent wall setup to the effective wall constant value for the Inconel vessel (rightmost value corresponding to  $\tau_{w,a}= 3\ \text{ms}$). For the transparent wall limit, SpeCyl predicts the $(m=2,n=1)$ RWM growth rate as $\gamma \sim 46\ \text{s}^{-1}$, while for the realistic $\tau_{w,a}$ it gives $\gamma \sim 35\ \text{s}^{-1}$. As a reference, the value of $\gamma = 47\ \text{s}^{-1}$ was produced by RFXlocking simulations\cite{Zanca2012Feedback}, which was shown to be in good agreement with the experimental observations for RFX-mod, giving $\gamma=44\ \text{s}^{-1}$, as already mentioned. A similar trend for varying $\tau_{w,a}$ is observed in the other two curves for $\tau_{w,b}=115\ \text{ms},\ \tau_{w,b}=125\ \text{ms}$, although with lower growth rates consistent with the longer time constants of the stabilizing shell.

Overall, the natural growth rates computed by SpeCyl fall in the correct order of magnitude as predicted by RFXlocking and observed in the target RFX-mod experiments. The best agreement is found for the nominal value of $\tau_{w,b}$ (as in RFXlocking) and in the transparent limit of the inner wall. On the other hand, when the inner wall is included with the nominal time constant as in RFXlocking, the resulting growth rate is $25\%$ smaller than what the latter predicted. However, it must be kept in mind that the SpeCyl result must be regarded as conservative because the inner wall is placed at the plasma boundary instead of being located at a finite distance as in RFXlocking, in which, due to the presence of graphite tiles, there is a $3\ \text{cm}$ gap between the plasma edge and the inner wall location. In addition, RFXlocking neglects plasma inertia, whereas SpeCyl accounts for it. Therefore, the outcome of this second and more realistic verification exercise of the new SpeCyl boundary module can be said to be substantially satisfactory and to confirm the improved predictive capabilities of the new code version. Future investigations will address a more in-depth validation with respect to RFX-mod observations and will also aim to deliver predictions for the upcoming RFX-mod2 device. %For this purpose, an appropriate resistivity value will be adopted for the first wall in contact with the plasma composed of carbon tiles mounted on a suitable support, to replace the Inconel vessel of RFX-mod. Moving outwards in vacuum and very close to the first wall  ($b/a=1.04$), the second wall will be modelled by the copper Passive Stabilizing Shell. The Vacuum Tight Support Structure (VTSS) will defining the double coaxial vacuum region, enclosing the previous elements.  

\section{Summary and final outlook}\label{sec:section9}

This paper presents a fundamental improvement of the SpeCyl boundary module: a double shell/double vacuum treatment for the magnetic field was derived and self-consistently merged with the pre-existing free flow treatment at the plasma edge \cite{Spinicci2023_MHD}, enabling the simulation of RWMs. This task required the numerical implementation of new equations in the boundary module, addressing a double wall and a double vacuum dynamics compliant with the spectral and nonlinear nature of SpeCyl. A successful extensive verification against Shafranov's analytical theory was performed for the $(m=2,n=1)$ resistive wall and external kink modes, based on a set of different equilibrium configurations characterized by uniform current profiles. Next, flexibility in the choice of the wall positions and of the wall time constants was fully exploited for the second satisfactory benchmark study against RFX-mod tokamak modelling, which proved the abilities of SpeCyl towards realistic experimental modelling and growth rates prediction. The two verifications confirmed the predictive capabilities of the numerical tool in the case of a circular section large aspect ratio tokamak and opened the path for the further use of SpeCyl for physics-based studies focused on the RFX-mod\cite{Piovesan2013,Peruzzo17} and RFX-mod2\cite{Peruzzo18,marrelli2024rfxmod2} machines, in both RFP and tokamak configurations. %The implementation just presented is fully self consistent and produces no numerical artifacts, even with high plasma density and conductivity at numerical boundary. This trait is still missing in most nonlinear MHD codes, despite being particularly valuable in modelling transient limited configurations in the tokamak and routine operation in the RFP, as detailed in \cite{Cappello2024Impact}.
In summary, the magneto-fluid boundary conditions are now addressed in the SpeCyl code with high-fidelity and self-consistency, enabling a reliable estimation of growth rates for free boundary instability scenarios. These features (similarly to the case of VDEs\cite{NIMROD,NIMROD_Bunkers2020,Jones2026PIXIE3D}), are crucial for the simulation of RWMs, ubiquitous and dangerous modes arising from the unavoidably finitely conducting structures enclosing the plasma.

%which are expected relevant instabilities for magnetic fusion devices due to the mandatory presence of conductive/resistive electromagnetic structures in the plasma surrounding.
%The experimental set-up characterizing the RFX-mod and RFX-mod2 machines defines a complex electromagnetic boundary, which in principle requires a detailed model including magnetic pick-up coils and feedback coils, in addition to the $3$D conducting walls.
%The improved BCs are believed to be a specific feature of SpeCyl: other MHD codes such as DEBS \cite{Paccagnella2007_RFXmod}, NIMROD \cite{NIMROD,NIMROD_BCs}, XTOR-2F \cite{XTOR-2F} and M$3$D \cite{M3D} implement a single resistive shell being a finite conductivity physical wall, while the code JOREK is coupled with the linear stability code  STARWALL \cite{JOREK-STARWALL,McAdams2014_Tokamak} to account for the electromagnetic boundary response. These codes adopt a pseudo-vacuum approach\cite{Cappello2024Impact} to describe the plasma edge, which is consistent with the natural presence of a separatrix and a scrape-off layer (SOL). Consequently, they enforce either a Dirichlet null radial flow velocity condition (no-slip condition) or a Neumann condition in correspondence with the physical resistive wall. An analogous numerical treatment was shown to be unfit for SpeCyl \cite{Cappello2024Impact}: this is the reason why the $3$D normal flow velocity treatment was preserved in this updated version of SpeCyl.

This study is part of a wider effort\cite{NatureRFX,Veranda_2017} aimed at improving the predictive capability of the SpeCyl code with a focus on both the tokamak and specifically the RFP configuration\cite{Cap08,Esc14}, which is actively investigated at Consorzio RFX (Padua, Italy). The coupling between internal MHD dynamics and a realistic magnetic front-end in the RFP configuration will be the subject of future investigations concerning the magnetic feedback control of the experiments. Adopting the double shell implementation as a basis, the further inclusion of feedback control coils is believed to be important for numerical studies on RFX-mod and RFX-mod2\cite{Marrelli19, Marrelli2026RFXmod2}, also considering the RFP interest in the improved confinement Quasi Single Helicity (QSH) states\cite{NatureRFX,PoP2010,Bonfiglio_2013}. These are self-organized states characterized by a large scale helical structure, with the dominance of one saturated mode over the rest of the spectrum. Until now, numerical investigations of QSH states have relied on the enforcement of Magnetic Perturbations (MPs) at the plasma edge: these externally set perturbations confirmed the possibility of inducing the selected helical state, driving the QSH plasma rearrangement\cite{Bonfiglio_2013}, and were able to sustain the achieved RFP configuration\cite{Paccagnella2007_RFXmod} on time scales longer than the penetration time of passive resistive shells. In addition, MPs were proven to be essential to mitigate sawtooth activity in MHD tokamak modelling\cite{Bonfiglio2017Sawtooth}. In general, the aim of these MPs was to mimic a modal radial magnetic field\cite{Bonfiglio_2013} at the plasma edge (which in real experiments is produced by control coils): the double shell implementation discussed in this work could make this picture even more realistic, since the second resistive shell located in the double coaxial vacuum (and the planned feedback control coils) is expected to naturally and self-consistently sustain the radial magnetic field modes at the plasma boundary. Moreover, the double shell implementation will improve the estimation of the expected MHD perturbation levels\cite{Marrelli19} at the RFX-mod2 boundary, which is crucial for a quantification of plasma-wall interaction determining the attainable high current plasma regimes. To conclude, the new implementation of SpeCyl achieved with this work allows for further physics-based studies aiming to closely reproduce the experimental settings. 

\section*{Author Contributions}
\textbf{Francesco Galeazzi}: Conceptualization (lead); Software (lead); Formal analysis (lead); Investigation (lead); Methodology (lead); Visualization (lead); Writing – original draft (lead); Writing – review \& editing (lead). 
\textbf{Daniele Bonfiglio}: Conceptualization (lead); Writing – original draft (equal); Writing – review \& editing (lead); Formal analysis (equal); Investigation (equal); Methodology (lead).
\textbf{Luca Spinicci}: Conceptualization (equal); Writing – review \& editing (lead); Formal analysis (equal); Investigation (equal); Methodology (equal).
\textbf{Susanna Cappello}: Conceptualization (equal); Writing – review \& editing (equal); Investigation (equal); Methodology (equal); Supervision (lead).

\begin{acknowledgments}
The authors would like to thank the colleagues G. Marchiori and P. Zanca for helpful discussions and valuable suggestions during the development of this paper. This work has been partially supported by the Program Implementation Agreement PRN MASE - ENEA of 2024/07/12 for activities set out in Art.3 c1d D.M. n. 386 2023/11/17 and Consorzio RFX. This work has been partially funded by the EPSRC Fusion Grant 2022/27 [grant number EP/W006839/1]. 

\end{acknowledgments}

\section*{Declaration of interest}
The authors have no conflict of interest to disclose.

\section*{Data availability statement}
Data supporting the findings of this study were obtained using SpeCyl (Consorzio RFX non-public code) and are stored on a private computing server. The data are available from the corresponding author upon request, subject to authorization from Consorzio RFX.

\appendix

\section{Double vacuum coefficients}\label{app:appB}
% Contenuto Appendice B
This section provides details about the magnetic field in the double vacuum region surrounding the cylindrical plasma. In addition, the expression of the double vacuum coefficients employed throughout the investigation is shown. In the vacuum surrounding the plasma, there is no charge or current density: the (quasi) magneto-static problem descends from the Maxwell equations assuming $\rho=0, \boldsymbol{J}=0$. The irrotational character of the magnetic field is met if each Fourier mode of the magnetic field satisfies
\begin{equation}\label{eq:def_pot}
\boldsymbol{B}^{m,n}= \nabla\varphi^{m,n}, 
\end{equation}
where $\varphi^{ m,n}$ is a scalar modal potential of shape
\begin{equation} \label{eq:phi_qs}   
\varphi^{ m,n}(r, \vartheta, z) = \phi^{m,n}(r)\,\exp(im\vartheta - inz/R_0).
\end{equation}
Using the solenoidal condition, the Laplace equation $\Delta\varphi^{m,n}=0$ is obtained; its version in cylindrical coordinates gives the modified Bessel equation for $\phi^{m,n}$:
%(using the change of variable $\tilde{r}=k_nr$):
\begin{equation} \label{eq:modified_bessel}  
r^2\dfrac{\d^2\phi^{m,n}(r)}{\d r^2} + r\dfrac{\d \phi^{m,n}(r)}{\d r} -\Big(m^2 +\dfrac{n^2}{R_0^2}r^2\Big)\,\phi^{m,n}(r) = 0.
\end{equation}
The potential then has the following expression, when $n\neq0$, and taking advantage of the subscript $l=1,2$ to address the vacuum region for which the expression holds:
\begin{equation} \label{eq:pot}  
\varphi_{l}^{m,n} = \Big[\mathcal{A}_lI_m\Big(\dfrac{|n|r}{R_0}\Big)+\mathcal{B}_lK_m\Big(\dfrac{|n|r}{R_0}\Big)\Big]\exp(im\vartheta -inz/R_0).
\end{equation}
The constants $\mathcal{A}_l,\mathcal{B}_l$ are determined by suitable boundary conditions, and $I_m,K_m$ are the modified Bessel functions\cite{Watson1966} of order $m$. The most general shape of this potential is introduced in the next lines, while a discussion of special cases (either the case of large aspect ratio equivalent to $n=0$, and the axisymmetric case $(m=0,n=0)$) can be found in Ref.\cite{Spinicci2023_MHD}. The general $n\neq0$ double-vacuum expressions for the magnetic field components are obtained by solving two BVPs accounting for the magnetic behaviour in correspondence of the vacuum domain edges: for a perfect conductor, the normal component of $\boldsymbol{B}$ is null, and so $B_r^{m,n}=0$. % because the normal component of $\boldsymbol{B}$ is requested to be continuous. 
For resistive materials, the sustained $B_r^{m,n}$ can be different from zero. In general, the tangential components of $\boldsymbol{B}$ are free but must meet the solenoidal condition. These physical conditions are related to $\varphi^{m,n}_l$ using Equation~\eqref{eq:def_pot}, which implies $B_{r,l}^{ m,n}= \partial\varphi^{m,n}_l/\partial r$. The first BVP to determine $\mathcal{A}_1,\mathcal{B}_1$ is solved in the vacuum region for which $a\leq r \leq b$, assuming two finite values for $B_r^{m,n}$ at the edges. These are $B_r^{m,n}(a),B_r^{ m,n}(b)$, determined according to the dynamics of the corresponding thin shell described by Equation~\eqref{eq:evol_Br}. Such fields are the computational unknowns managed and evolved by SpeCyl, resulting fundamental in defining the equations of the renewed boundary module, as presented in Section~\ref{sec:section4}. The second BVP to determine $\mathcal{A}_2,\mathcal{B}_2$ is solved in the vacuum region for which $b \leq r \leq c$ and imposes % $B_b^{m,n}= 
$B_r^{ m,n}(b)$, $B_r^{m,n}(c)= 0$ as boundary values. The ideal wall radius, i.e. $r_w=c$, is typically set to a large value, so that the presence of such a wall does not affect the dynamics and is only used to close the system of equations. However, if needed, it can also be placed at a finite distance to the plasma and other conducting elements, for instance to represent a thick/quasi-ideal structure. 

The modal values of $\mathcal{A}_1,\mathcal{B}_1,\mathcal{A}_2,\mathcal{B}_2$ are reported below, built as combinations of the derivatives of modified Bessel functions (denoted $I_m^{'}, K_m^{'}$) and highlighting the role of shell-fields. Each of the two vacuums is addressed by the subscript $l=1,2$:
\begin{equation}\label{eq:coeffs_1}  
\begin{split}
    &\mathcal{A}_1^{m,n}=\alpha_{1a}^{mn}\ B_r^{ m,n}(a) + \alpha_{1b}^{mn}\ B_r^{ m,n}(b),\\ 
    &\mathcal{B}_1^{m,n}=\beta_{1a}^{mn}\ B_r^{ m,n}(a) + \beta_{1b}^{mn}\ B_r^{ m,n}(b),\\
    &\mathcal{A}_2^{m,n}=\alpha_{2b}^{mn}\ B_r^{ m,n}(b),\\ 
    &\mathcal{B}_2^{m,n}=\beta_{2b}^{mn}\ B_r^{ m,n}(b).\\
\end{split}
\end{equation}
These constants are constructed according to:
\begin{equation}\label{eq:coeffs_0}
  \begin{split}
   &\alpha_{1a}^{mn} = \dfrac{K'_{m}\Bigl(\dfrac{|n|b}{R_0}\Bigr)}{I'_{m}\Bigl(\dfrac{|n|a}{R_0}\Bigr)K'_{m}\Bigl(\dfrac{|n|b}{R_0}\Bigr) - I'_{m}\Big(\dfrac{|n|b}{R_0}\Big)K'_{m}\Big(\dfrac{|n|a}{R_0}\Big)},\\
   &\alpha_{1b}^{mn} = \dfrac{-K'_{m}\Bigl(\dfrac{|n|a}{R_0}\Bigr)}{I'_{m}\Bigl(\dfrac{|n|a}{R_0}\Bigr)K'_{m}\Bigl(\dfrac{|n|b}{R_0}\Bigr) - I'_{m}\Big(\dfrac{|n|b}{R_0}\Big)K'_{m}\Big(\dfrac{|n|a}{R_0}\Big)},\\
   &\beta_{1a}^{mn} = \dfrac{I'_{m}\Bigl(\dfrac{|n|b}{R_0}\Bigr)}{I'_{m}\Big(\dfrac{|n|b}{R_0}\Big)K'_{m}\Bigl(\dfrac{|n|a}{R_0}\Bigr) - I'_{m}\Big(\dfrac{|n|a}{R_0}\Big)K'_{m}\Big(\dfrac{|n|b}{R_0}\Big)},\\
   &\beta_{1b}^{mn} = \dfrac{-I'_{m}\Bigl(\dfrac{|n|a}{R_0}\Bigr)}{I'_{m}\Big(\dfrac{|n|b}{R_0}\Big)K'_{m}\Bigl(\dfrac{|n|a}{R_0}\Bigr) - I'_{m}\Big(\dfrac{|n|a}{R_0}\Big)K'_{m}\Big(\dfrac{|n|b}{R_0}\Big)},\\
   &\alpha_{2b}^{mn} = \dfrac{-K'_{m}\Bigl(\dfrac{|n|c}{R_0}\Bigr)}{I'_{m}\Big(\dfrac{|n|c}{R_0}\Big)K'_{m}\Bigl(\dfrac{|n|b}{R_0}\Bigr) - I'_{m}\Big(\dfrac{|n|b}{R_0}\Big)K'_{m}\Big(\dfrac{|n|c}{R_0}\Big)},\\
   &\beta_{2b}^{mn} = \dfrac{I'_{m}\Bigl(\dfrac{|n|c}{R_0}\Bigr)}{I'_{m}\Big(\dfrac{|n|c}{R_0}\Big)K'_{m}\Bigl(\dfrac{|n|b}{R_0}\Bigr) - I'_{m}\Big(\dfrac{|n|b}{R_0}\Big)K'_{m}\Big(\dfrac{|n|c}{R_0}\Big)}.\\
   \end{split}
\end{equation}
Taking advantage of the latter expressions, the radial magnetic field modes result as follows, for the first and second vacuums (see Equation~\eqref{eq:Br} in Section~\ref{sec:section3}):
\begin{equation}\label{eq:expBr1}
B_{r,1}^{m,n} =\Big[\mathcal{A}_1^{m,n}I_m^{'}\Big(\dfrac{|n|r}{R_0}\Big) + \mathcal{B}_1^{m,n}K_m'\Big(\dfrac{|n|r}{R_0}\big)\Big]\exp(im\vartheta - inz/R_0),
\end{equation}
\begin{equation}\label{eq:expBr2}
B_{r,2}^{m,n} =\Big[\mathcal{A}_2^{m,n}I_m^{'}\Big(\dfrac{|n|r}{R_0}\Big) + \mathcal{B}_2^{m,n}K_m'\Big(\dfrac{|n|r}{R_0}\Big)\Big]\exp(im\vartheta - inz/R_0).
\end{equation}
Along tangential directions ($\vartheta,z$), Equation~\eqref{eq:def_pot} combined with the assumed periodic character causes algebraic conditions dependent on the shape of $\varphi^{m,n}_l$ to arise. These yield, taking advantage of Equation~\eqref{eq:pot}:
\begin{equation}\label{eq:btbr}
B_{\vartheta,l}^{m,n}=-i\dfrac{m R_0}{|n|r}\Big[\mathcal{A}_l^{m,n}I_{m}\Big(\dfrac{|n|r}{R_0}\Big) + \mathcal{B}_l^{m,n}K_{m}\Big(\dfrac{|n|r}{R_0}\Big)\Big]\exp(im\vartheta - inz/R_0),
\end{equation}
and
\begin{equation}\label{eq:bzbr}
B_{z,l}^{m,n}=i\dfrac{n}{|n|}\Big[\mathcal{A}_l^{m,n}I_{m}\Big(\dfrac{|n|r}{R_0}\Big) + \mathcal{B}_l^{m,n}K_{m}\Big(\dfrac{|n|r}{R_0}\Big)\Big] \exp(im\vartheta - inz/R_0).
\end{equation}
Equations~\eqref{eq:btbr},~\eqref{eq:bzbr} are rewritten to highlight the role of $B_r^{m,n}(a), B_r^{m,n}(b)$ and evaluated for $r=a$, to comply with their use in 
Equations~\eqref{eq:third_disc},~\eqref{eq:fourth_disc} of the renewed boundary module aiming at enforcing the continuity of tangential electric fields for the thin shell in $r=a$:
\begin{equation}\label{eq:tancomp}  
\begin{split}
&B_{\vartheta,l}^{m,n}=i\Theta_{laa}^{m,n}\ B_r^{m,n}(a) + i\Theta_{lab}^{m,n}\ B_r^{m,n}(b),\\
   &B_{z,l}^{m,n}=iZ_{laa}^{m,n}\ B_r^{m,n}(a) + iZ_{lab}^{m,n}\ B_r^{m,n}(b).\\
\end{split}
\end{equation}
These, in turn, require the definition of (again $l=1,2$):
\begin{equation}\label{eq:TZ}
\begin{split}
 &\Theta_{laa}^{m,n} = \dfrac{mR_0}{|n|a}\Big[\alpha_{la}^{mn}\ I_{m}\Big(\dfrac{|n|a}{R_0}\Big) + \beta_{la}^{mn}\ K_{m}\Big(\dfrac{|n|a}{R_0}\Big) \Big],\\
 &\Theta_{lab}^{m,n} = \dfrac{mR_0}{|n|a}\Big[\alpha_{lb}^{mn}\ I_{m}\Big(\dfrac{|n|a}{R_0}\Big) + \beta_{lb}^{mn}\ K_{m}\Big(\dfrac{|n|a}{R_0}\Big) \Big],\\
 &Z_{laa}^{m,n} = \dfrac{n}{|n|}\Big[\alpha_{la}^{mn}\ I_{m}\Big(\dfrac{|n|a}{R_0}\Big) + \beta_{la}^{mn}\ K_{m}\Big(\dfrac{|n|a}{R_0}\Big) \Big],\\
 & Z_{lab}^{m,n} = \dfrac{n}{|n|}\Big[\alpha_{lb}^{mn}\ I_{m}\Big(\dfrac{|n|a}{R_0}\Big) + \beta_{lb}^{mn}\ K_{m}\Big(\dfrac{|n|a}{R_0}\Big) \Big].\\
\end{split}
\end{equation}
This appendix ends by presenting the double vacuum coefficients requested by the evolution of the general harmonic $(n\neq0)$ of the radial magnetic field in correspondence with the two thin shells in $r=a,b$ (Equations~\eqref{eq:first_disc},~\eqref{eq:second_disc}). Their expression is built on the expressions discussed so far and reads:
\begin{equation}\label{eq:set_Cljk}
\begin{split}
 &\mathcal{C}_{1aa}^{m,n} = \dfrac{R_0}{|n|a}\Big(m^2+\dfrac{n^2a^2}{R_0^2}\Big)\Big[\alpha_{1a}^{mn}\ I_{m}\Big(\dfrac{|n|a}{R_0}\Big) + \beta_{1a}^{mn}\ K_{m}\Big(\dfrac{|n|a}{R_0}\Big)\Big],\\
 &\mathcal{C}_{1ab}^{m,n} = \dfrac{R_0}{|n|a}\Big(m^2+\dfrac{n^2a^2}{R_0^2}\Big)\Big[\alpha_{1b}^{mn}\ I_{m}\Big(\dfrac{|n|a}{R_0}\Big) + \beta_{1b}^{mn}\ K_{m}\Big(\dfrac{|n|a}{R_0}\Big) \Big],\\
 &\mathcal{C}_{1ba}^{m,n} = \dfrac{R_0}{|n|b}\Big(m^2+\dfrac{n^2b^2}{R_0^2}\Big)\Big[\alpha_{1a}^{mn}\ I_{m}\Big(\dfrac{|n|b}{R_0}\Big) + \beta_{1a}^{mn}\ K_{m}\Big(\dfrac{|n|b}{R_0}\Big) \Big],\\
 &\mathcal{C}_{1bb}^{m,n} = \dfrac{R_0}{|n|b}\Big(m^2+\dfrac{n^2b^2}{R_0^2}\Big)\Big[\alpha_{1b}^{mn}\ I_{m}\Big(\dfrac{|n|b}{R_0}\Big) + \beta_{1b}^{mn}\ K_{m}\Big(\dfrac{|n|b}{R_0}\Big) \Big],\\
 &\mathcal{C}_{2bb}^{m,n} = \dfrac{R_0}{|n|b}\Big(m^2+\dfrac{n^2b^2}{R_0^2}\Big)\Big[\alpha_{2b}^{mn}\ I_{m}\Big(\dfrac{|n|b}{R_0}\Big) + \beta_{2b}^{mn}\ K_{m}\Big(\dfrac{|n|b}{R_0}\Big) \Big].\\
\end{split}
\end{equation}
For $n=0$, i.e. the large aspect ratio case, SpeCyl adopts suitable coefficients in which Bessel functions are replaced by polynomials of order $m$.

\section{MHD stability in the straight tokamak limit}\label{app:appC}% Contenuto Appendice C

The MHD linear stability analysis, employed for the derivation of the theoretical growth rates mentioned in this work, is here reviewed
in the limit of a vanishing inverse aspect ratio tokamak with circular cross section. The study departs from a static equilibrium ($\boldsymbol{v}_0=0$) and adopts a perturbative perspective, having the poloidal flux function $\psi(r,\vartheta,z;t)$ and the instability growth rate $\gamma$ as the main unknowns of the problem. Equilibrium quantities are denoted by the subscript \hh{$0$}, and perturbed quantities are denoted by a \hh{$\delta$} placed before their respective symbols. The starting MHD equations\cite{Biskamp} are the linearized momentum equation (which in principle considers the pressure contribution) and the linearized induction equation (the second is generated by a suitable combination of Equations~\eqref{eq:ind} -~\eqref{eq:divlessB}), as follows: 
\begin{equation}\label{eq:linearized_mom}
    \rho_0\dfrac{\partial\delta \boldsymbol{v}}{\partial t}  = -\nabla \delta p + \rho_0\nu_0\Delta\delta \boldsymbol{v}+ \delta\boldsymbol{j} \times \boldsymbol{B}_0 + \boldsymbol{J}_0 \times \delta\boldsymbol{b},
\end{equation}
\begin{equation}\label{eq:linearized_mag}
    \dfrac{\partial\delta\boldsymbol{b}}{\partial t} = \nabla\times(\delta\boldsymbol{v}\times\boldsymbol{B}_0)+\dfrac{\eta}{\mu_0}\Delta\delta\boldsymbol{b} .
\end{equation}
In the derivation of these equations from the basic MHD plasma modelling, the density field has been assumed constant, in compliance with the approach followed by SpeCyl\cite{Cap_Bisk_96}. The perturbed velocity and magnetic fields ($\delta\boldsymbol{v}$ and $ \delta\boldsymbol{b}$, respectively) are as: $\delta \boldsymbol{v} = \nabla\times(\varphi\hat{\boldsymbol{u_z}}),\ \delta \boldsymbol{b} = \nabla\times(\psi\hat{\boldsymbol{u_z}})$, where $\hat{\boldsymbol{u_z}}$ is the axial (or toroidal) unit vector and $\psi$ and $\varphi$ are the poloidal flux function and the stream function, respectively. The main unknown is $\psi$, because the ideal version ($\eta=0)$ of Equation~\eqref{eq:linearized_mag} produces the relation $\varphi=\varphi(\psi)$. The modal expression for $\psi$ is written as follows, so that Equations~\eqref{eq:linearized_mom},~\eqref{eq:linearized_mag} are Fourier transformed in the periodic coordinates $\vartheta,z$:
%(without writing its modal character, i.e. writing $\psi$ in place of $\psi^{ m,n}$):
\begin{equation} \label{eq:def_psi}
\psi^{m,n}(r,\vartheta,z; t) = \Psi^{m,n}(r)\,\exp(im\vartheta -inz/R_0)\exp(\gamma t).
\end{equation}
The two key elements in Equation~\eqref{eq:def_psi} are the mode eigenfunction $\Psi^{m,n}(r)$ and the mode growth rate $\gamma\in\mathbb{C}$. Two important relations involving $\psi^{m,n}$ are:
\begin{align}
&\xi_r^{m,n}=\psi^{m,n}/F(r),\label{eq:link_1}\\
&B_r^{m,n}= \dfrac{im}{r}\psi^{m,n}.\label{eq:link_2}
\end{align}
The first arises from the indefinite partial time-integrated ideal perturbed evolution equation for $\boldsymbol{B}$, once set $\delta\boldsymbol{v}=\partial\boldsymbol{\xi}/\partial t$ and where $\boldsymbol{\xi}$ is the plasma displacement. The factor $F(r)$ is a function dependent on the mode and equilibrium considered. Its expression reads
\begin{equation}\label{eq:factor_F}
F(r)=B_{\vartheta}^{0,0}-(nr/mR_0)B_{z}^{0,0},
\end{equation}
in which the superscript $(0,0)$ denotes equilibrium fields. The second condition arises directly from the radial projection of $\boldsymbol{B}^{m,n} = \nabla \times (\psi \hat{\boldsymbol{u}}_z)$, in which the perturbed radial magnetic field $\delta b_r$ is rewritten as $B_r^{m,n}$, in accordance with the SpeCyl notation. From now on, the modal indices $(m,n)$ appearing in $\psi^{m,n}, \Psi^{m,n}$ are removed to access a more compact notation.\\
The Laplace operator acting on the scalar field $\psi$, decomposed into its orthogonal part (the one associated with coordinates $r,\vartheta$) and into its parallel part (associated with coordinate $z$), reads
\begin{equation} \label{eq:def_lap}  
\Delta \psi =
\underbrace{
\frac{1}{r}\frac{\partial}{\partial r}\Bigl(r\frac{\partial\psi}{\partial r}\Bigr)
+ \frac{1}{r^2}\frac{\partial^2\psi}{\partial \vartheta^2}
}_{\Delta_\perp \psi}
\quad
+
\quad
\underbrace{
\frac{\partial^2\psi}{\partial z^2}
}_{\Delta_\parallel \psi}.
\end{equation}
The perpendicular part $\Delta_\perp \psi$ enters the main theoretical tool adopted to derive the dispersion relation, whereas the parallel term vanishes in the large aspect ratio limit. The torque balance equation (also adopted in Ref.\cite{McAdams2014_Tokamak}) describes the electromagnetic torques acting on the unstable plasma column and is now derived in the ideal limit of MHD, i.e., assuming vanishing plasma resistivity $\eta$ and dynamic viscosity $\mu$. An equation for $\psi$ is obtained by applying the curl to the perturbed momentum balance equation (Equation~\eqref{eq:linearized_mom}) and isolating its $z$ component, after some manipulations:
\begin{equation}\label{eq:torque_balance}  
\begin{split}
    \Delta_\perp\psi - \frac{\mu_0 m}{B_\vartheta(m-nq)}{r}\frac{\partial J_z}{\partial r}\psi &=-\gamma^2\dfrac{\mu_0r}{B_\vartheta(m-nq)}\\
    &\cdot\Big[\rho_0\Delta_\perp\Big(\frac{r}{B_\vartheta(m-nq)}\psi\Big)
    + \frac{\partial \rho_0}{\partial r}\frac{\partial}{\partial r}\Big(\frac{r}{B_\vartheta(m-nq)}\psi\Big)\Big].
\end{split}
\end{equation}
The previous equation involves the safety factor $q$ for a circular cross section toroidal plasma, as expressed by Equation~\eqref{eq:safety_factor}.
%Its expression is recalled:
%\begin{equation}\label{eq:safetyf}
%  q(r) = \dfrac{rB_z}{R_0B_\vartheta}, \end{equation}
%to make evident that the safety factor allows the features of the configuration of interest (other than mode-dependent ones) to enter the picture. 
Through $q(r)$, the specific version of the torque balance equation (Equation~\eqref{eq:torque_balance}) will depend on the equilibrium profiles adopted for 
$B_\vartheta(r)$, $J_z$ (or equivalently $B_z$), and $\rho_0$. 
%The considerations concerning the singular character of Equation~\eqref{eq:torque_balance} are postponed to the next section. Notice that here $\rho_0$ denotes the equilibrium profile for the density adopted in the perturbative investigation, which decomposes the plasma density as $\rho=\rho_0 +\delta\rho$. In the following section, a slightly different notation for $\rho$ is adopted. 
The next step consists of integrating the torque balance equation over the radial coordinate, after having assumed a specific shape for the above mentioned profiles. The integration makes then evident that the only relevant quantities entering Equation~\eqref{eq:def_psi} are $\Psi(r)$ and $\gamma$, as anticipated. These results are basically equivalent to those achievable with the Energy Principle \cite{Freid,Freidberg_plasma_physics_fusion_energy}, whose main unknown is the plasma displacement $\boldsymbol{\xi}$. 
This second approach is only briefly outlined to present some of its basic features necessary to define the characteristic features of the Shafranov's FCM: in particular, it will yield the shape for $\Psi$. Plasma displacement $\boldsymbol{\xi}$ is Fourier decomposed as $\boldsymbol{\xi}=\sum_{m,n}\boldsymbol{\xi}^{m,n}$, in which the plasma modal displacement is $\boldsymbol{\xi}^{m,n} = \xi_r^{m,n}\hat{\boldsymbol{u}}_r + \boldsymbol{\xi}_t^{ m,n}$. The equivalence of the two methods is discussed in Ref.\cite{Cappello2024Impact} and is based on the relation already presented, i.e. Equation~\eqref{eq:link_1}.
%The physical meaning subtended to this relation is highlighted in the end of the Appendix~\ref{app:appB}, together with its expression.
The result presented so far will now become particular, once the profiles $J_z, \rho_0$ are given a specific shape.

\section{Analytic details on the Shafranov's FCM}\label{app:appD} % Contenuto Appendice D
The Shafranov's flat current model\cite{Shafranov1970} assumes for the equilibrium axial current density and mass density a sharp step profile in correspondence to the plasma edge, expressed as
$\boldsymbol{J} = J_0 H(a - r)\hat{\boldsymbol{u}}_z ,\ \rho  = \rho_0 H(a - r)$. The static Ampère-Maxwell law projected along $z$ allows writing the poloidal component of $\boldsymbol{B}$ at equilibrium:
\begin{equation} \label{eq:B_theta_full}   
 B_\vartheta(r)= 
 \begin{cases}
 \frac{1}{2}\mu_0 J_0 r  & \text{if $r \leq a$,} \\
 \frac{1}{2}\mu_0 J_0\frac{a^2}{r}  & \text{if $r > a$}.
 \end{cases}
\end{equation}
The axial magnetic field $B_z$ is assumed to be uniform, i.e. $B_z(r)=B_0$. These results allow computing the expression of the safety factor, where $q_0=2B_0/\mu_0J_0R_0$:
\begin{equation} \label{eq:sf_full} 
 q(r)= 
 \begin{cases}
 q_0  & \text{if $r \leq a$,} \\
 q_0 \frac{r^2}{a^2}  & \text{if $r > a$}.
 \end{cases}
\end{equation}
These two profiles (represented in Figure~\ref{fig:eq_profiles}) are crucial to characterize the torque balance, i.e. Equation~\eqref{eq:torque_balance}. %In addition, the consequent magnetic shear profile, whose definition is $s(r)= -r/q(r)\cdot \d q(r)/ \d r$, has a profound impact on the study of tearing modes:
%\begin{equation} \label{eq:shear} 
% s(r)= 
% \begin{cases}
% 0  & \text{if %$r \leq a$,} \\
% -2 & \text{if $r > a$}.
% \end{cases}
%\end{equation}
%Then, due to the null value, the flat current assumption eliminates the stabilization mechanism provided by the magnetic field line bending. The local ordering of the resistive layer equations is modified for the tearing mode investigation: the mode can result with an enhanced global structure because the resonance is not locally isolated. In practice, the classical two-region approach with an inner resistive layer \cite{furth1963finite} is not directly applicable because the classical tearing mode scaling is violated. In the present work, both $\eta,\mu$ are sufficiently low (high values for dimensionless $\mathit{S}$ and $\mathit{M}$) that the dynamics is dominated by ideal kink-type instabilities, eventually transitioning to RWMs.
Once substituted and having introduced a suitable numerical factor proportional to $q_0^2\tau_A^2$, the torque balance equation reads
\begin{equation}\label{eq:torque_balance_cfm}  
\begin{split}
    &\Delta_\perp\psi - \frac{2m}{J_0(m-nq_0)r}\frac{\partial J_z}{\partial r}\psi = \frac{-\gamma^2\tau_A^2/\epsilon^2q_0^2}{(m-nq_0)^2}\cdot\Bigl[\dfrac{\rho}{\rho_0}\Delta_\perp\psi+ \frac{\partial}{\partial r}\Bigl(\frac{\rho}{\rho_0}\Bigr)\dfrac{\partial\psi}{\partial r}\Bigr].
\end{split}
\end{equation}
The latter is singular when a resonance is met, i.e., when $q(r_s) = m/n=q_0$. In the current investigation $q(r_s)=2$ and the resonance radius $r_s$ is externally localized with respect to the plasma.
%in all the simulated cases, with the only exception given by the last equilibrium considered in Figure~\ref{fig:scan_qa}. In such a case, the resonance radius is met in the whole plasma domain due to Shafranov's flat current model assumptions. 
Equation~\eqref{eq:torque_balance_cfm} holds for $r\leq a$ and is rewritten multiplicating by $r$ and eliminating the function $\psi$ from all terms because its radial partial derivative can be expressed proportionally to $\psi$, i.e. $\partial\psi/\partial r= \psi\Psi^{'}/\Psi$. The meaning of $\Psi(r), \Psi^{'}(r)$ is clear from Equation~\eqref{eq:def_psi}. The resulting equation is further integrated on the radial coordinate, as anticipated in the previous section, defining a suitable integration domain across $r=a$ using positively and negatively displaced radii addressed by means of $a^+,a^-$. This procedure is simplified by the presence of Dirac delta functions, generated by the partial radial derivatives of quantities dependent on the Heaviside step function: $\partial J_z/\partial r\sim\delta(a-r),\ \partial\rho/\partial r\sim\delta(a-r)$. The polynomial term arising in $\Delta_\perp\psi$ (the one associated with the partial derivative in $\vartheta$, see Equation~\eqref{eq:def_lap}) is neglected because its continuity in $r$ cancels the corresponding contribution after integration. By doing so, the right hand side of Equation~\eqref{eq:torque_balance_cfm} is rewritten as the partial radial derivative of a product containing a Heaviside step function, which then cancels out the corresponding contribution for $a^+$. The torque balance equation describes the electromagnetic torques $\boldsymbol{J}\times\boldsymbol{B}$ that develop when the equilibrium is perturbed, ending in a differential condition for $\Psi(r)$ that indeed describes the linearized modal-dependent dynamics of the unstable cylindrical plasma:
\begin{equation}\label{eq:dispersion_rel_1}
r\Big[\frac{\Psi^{'}(r)}{\Psi(r)}\Big]\Bigg\vert_{a} + \frac{2m}{(m-nq_0)} = \frac{\gamma^2\tau_A^2/\epsilon^2q_0^2}{(m-nq_0)^2}r\frac{\Psi^{'}(r)}{\Psi(r)}\Bigg\vert_{a^-}.
\end{equation}
The square brackets in the first term denote the evaluation across the radial position $r=a$, generating two terms (for $a^+,a^-$). The expected discontinuity of $\Psi$ is often referenced in the kink or tearing mode investigation literature\cite{McAdams2014_Tokamak, NIMROD_BCs} using the symbol $\Delta'$. The previous condition is used as the first boundary condition to solve a suitable BVP for $\Psi(r)$ in the vacuum surrounding the plasma, where it obeys $\Psi(r)=Ar^m + Br^{-m}$. The shape of this general solution is easily retrieved when solving Equation~\eqref{eq:torque_balance_cfm} in vacuum, i.e. for vanishing current $J_z$ and density $\rho$. The second boundary condition arises by integrating over $r$ the diffusion equation for the magnetic field perturbation in a thin shell at $r=r_w$, generated from Equation~\eqref{eq:linearized_mag} when the flow field is vanishing, i.e.:
\begin{equation}\label{eq:diffusion_pert_b}
\dfrac{\partial\delta\boldsymbol{b}}{\partial t} = \dfrac{\eta_w}{\mu_0}\Delta\delta\boldsymbol{b}\hspace{0.3cm}\rightarrow\hspace{0.3cm}r\Big[\dfrac{\Psi^{'}(r)}{\Psi(r)}\Big]\Bigg|_{r_w} = \gamma\tau_w.
\end{equation}
The solution of the BVP produces the dispersion relation of the mode, conveyed by Equation~\eqref{eq:rwm_full}, which accounts for both the plasma response at the MHD level (condition~\eqref{eq:dispersion_rel_1}) and the presence of the thin resistive wall in $r_w$ (described by condition~\eqref{eq:diffusion_pert_b}). Although only a single resistive shell has been included in the analytical model presented here, the discussion can be extended to multiple wall configurations\cite{Liu2008Analytical} and also to the direct inclusion of feedback coils. \\
The appendix ends by showing how Shafranov's flat current model assumptions are employed to predict the shape for the radial magnetic field modes, given all the elements introduced so far. To obtain $\Psi$ or $\psi$, the easiest route is to rely on results that can be derived starting from the final considerations of Appendix~\ref{app:appC}. By simple arguments departing from the Energy Principle\cite{Freid,Freidberg_plasma_physics_fusion_energy} and setting the analysis in the straight tokamak case for an incompressible plasma with negligible pressure gradients\cite{Cappello2024Impact}, the plasma modal displacement $\boldsymbol{\xi}^{m,n}$ results in the $r-\vartheta$ (poloidal) plane and reads:
\begin{equation} \label{eq:displacement}   
\boldsymbol{\xi}^{m,n}= \xi_{r,a}^{m,n}\Big(\dfrac{r}{a}\Big)^{m-1}(\hat{\boldsymbol{u}}_r + i\hat{\boldsymbol{u}}_\vartheta)\exp(im\vartheta-inz/R_0-i\omega t),
\end{equation}
where $\xi_{r,a}^{m,n}$ is the edge value of the radial displacement and $\omega$ plays a role similar to $\gamma$. For the mode $(m=2,n=1)$, the radial plasma displacement results in a linear function in $r$, because $\xi_r^{2,1}\sim(r/a)$. Given the same mode and taking advantage of the equilibrium profiles characterizing the FCM (Equation~\eqref{eq:B_theta_full}), the shape of the factor $F(r)$ results:

\begin{equation} \label{eq:factor_F_shaf}   
 F(r)= 
 \begin{cases}
 \frac{1}{2}\mu_0 J_0(1 - q_0/2)r & \text{if $r \leq a$,} \\
 \frac{1}{2}\mu_0 J_0(a^2/r- q_0r/2)  & \text{if $r > a$}. 
 \end{cases}
\end{equation}
Combining the last result with Equations~\eqref{eq:link_1},~\eqref{eq:link_2} produces the analytical shape for radial magnetic field mode $B_r^{2,1}$ in the plasma domain. It results in a linear function in $r$, as $\xi_r^{2,1}$ is for $r\leq a$. In fact:
\begin{equation} \label{eq:br_sfcm}
B_r^{2,1}=\dfrac{2i}{r}F(r)\xi_r^{2,1}\sim\frac{r}{a}.
\end{equation}
The analytical shape prediction for $B_r^{2,1}$ is used in Section~\ref{sec:section6}, where the comparison of the analytical radial profiles with the one obtained in a typical SpeCyl simulation is provided.

%--------------------------------
%BIBLIOGRAFIA
%\bibliographystyle{apsrev4-2}
\bibliographystyle{aipnum4-2}
\nocite{*}
\bibliography{ArtBib}% Produces the bibliography via BibTeX.

\end{document}